\input harvmac.tex

\let\includefigures=\iftrue
\let\useblackboard=\iftrue
\newfam\black

\includefigures
\message{If you do not have epsf.tex (to include figures),}
\message{change the option at the top of the tex file.}
\input epsf
\def\figin{\epsfcheck\figin}\def\figins{\epsfcheck\figins}
\def\epsfcheck{\ifx\epsfbox\UnDeFiNeD
\message{(NO epsf.tex, FIGURES WILL BE IGNORED)}
\gdef\figin##1{\vskip2in}\gdef\figins##1{\hskip.5in}
\else\message{(FIGURES WILL BE INCLUDED)}%
\gdef\figin##1{##1}\gdef\figins##1{##1}\fi}
\def\DefWarn#1{}
\def\figinsert{\goodbreak\midinsert}
\def\ifig#1#2#3{\DefWarn#1\xdef#1{fig.~\the\figno}
\writedef{#1\leftbracket fig.\noexpand~\the\figno}%
\figinsert\figin{\centerline{#3}}\medskip\centerline{\vbox{
\baselineskip12pt\advance\hsize by -1truein
\noindent\footnotefont{\bf Fig.~\the\figno:} #2}}
\bigskip\endinsert\global\advance\figno by1}
\else
\def\ifig#1#2#3{\xdef#1{fig.~\the\figno}
\writedef{#1\leftbracket fig.\noexpand~\the\figno}%
\global\advance\figno by1}
\fi

\def\doublefig#1#2#3#4{\DefWarn#1\xdef#1{fig.~\the\figno}
\writedef{#1\leftbracket fig.\noexpand~\the\figno}%
\figinsert\figin{\centerline{#3\hskip1.0cm#4}}\medskip\centerline{\vbox{
\baselineskip12pt\advance\hsize by -1truein
\noindent\footnotefont{\bf Fig.~\the\figno:} #2}}
\bigskip\endinsert\global\advance\figno by1}

\useblackboard
\message{If you do not have msbm (blackboard bold) fonts,}
\message{change the option at the top of the tex file.}
\font\blackboard=msbm10 scaled \magstep1
\font\blackboards=msbm7
\font\blackboardss=msbm5
\textfont\black=\blackboard
\scriptfont\black=\blackboards
\scriptscriptfont\black=\blackboardss

\else

\fi
%
\def\subsubsec#1{\bigskip\noindent{\it{#1}} \bigskip}
\def\yboxit#1#2{\vbox{\hrule height #1 \hbox{\vrule width #1
\vbox{#2}\vrule width #1 }\hrule height #1 }}
\def\fillbox#1{\hbox to #1{\vbox to #1{\vfil}\hfil}}
\def\ybox{{\lower 1.3pt \yboxit{0.4pt}{\fillbox{8pt}}\hskip-0.2pt}}
%
%
\def\ep{\epsilon}


\def\comments#1{}

\def\half{{1\over 2}}
\def\Tr{{{\rm Tr~ }}}
\def\tr{{\rm tr\ }}
\def\Re{{\rm Re\hskip0.1em}}

\def\bra#1{{\langle}#1|}
\def\ket#1{|#1\rangle}

\def\vev#1{\langle{#1}\rangle}

\def\CA{{\cal A}}

\def\CO{{\cal O}}

\def\CV{{\cal V}}


\def\II{\relax{I\kern-.10em I}}

\def\IZ{\relax\ifmmode\mathchoice
{\hbox{\cmss Z\kern-.4em Z}}{\hbox{\cmss Z\kern-.4em Z}}
{\lower.9pt\hbox{\cmsss Z\kern-.4em Z}}
{\lower1.2pt\hbox{\cmsss Z\kern-.4em Z}}
\else{\cmss Z\kern-.4emZ}\fi}
\def\IB{\relax{\rm I\kern-.18em B}}
\def\IC{{\relax\hbox{$\inbar\kern-.3em{\rm C}$}}}
\def\ID{\relax{\rm I\kern-.18em D}}
\def\IE{\relax{\rm I\kern-.18em E}}
\def\IF{\relax{\rm I\kern-.18em F}}
\def\IG{\relax\hbox{$\inbar\kern-.3em{\rm G}$}}
\def\IGa{\relax\hbox{${\rm I}\kern-.18em\Gamma$}}
\def\IH{\relax{\rm I\kern-.18em H}}
\def\II{\relax{\rm I\kern-.18em I}}
\def\IK{\relax{\rm I\kern-.18em K}}
\def\IP{\relax{\rm I\kern-.18em P}}

%

\def\inbar{\,\vrule height1.5ex width.4pt depth0pt}

\font\cmss=cmss10 
\def\IR{\relax{\rm I\kern-.18em R}}

\def\one {{\bf 1}}

%


%

\def\lp10{\ell_p^{10}}
\def\lp11{\ell_p^{11}}
\def\R11{R_{11}}

\def\frac#1#2{{#1 \over #2}}









\def\1dag{^{1\dagger}}
\def\2dag{^{2\dagger}}

\def\R#1#2#3{{{R_{#1}}^{#2}}_{#3}}

\def\ot{\otimes}


\def\Ga{\Gamma}
\def\eg{{\it e.g.}}
\def\ie{{\it i.e.}}

\hyphenation{Di-men-sion-al}



\def\fr{\frac}
\def\ra{\rightarrow}
\def\pa{\partial}
\def\bz{{\bar z}}
\def\al{\alpha}

\lref\NiarchosKW{
V.~Niarchos,
``Density of states and tachyons in open and closed string theory,''
JHEP {\bf 0106}, 048 (2001)
[arXiv:hep-th/0010154].
}

\lref\GaiottoRM{
D.~Gaiotto, N.~Itzhaki and L.~Rastelli,
``Closed strings as imaginary D-branes,''
arXiv:hep-th/0304192.
}

\lref\McGreevyKB{
J.~McGreevy and H.~Verlinde,
``Strings from Tachyons,''
arXiv:hep-th/0304224.
}

\lref\DO{
H.~Dorn and H.~J.~Otto,
``Two and three point functions in Liouville theory,''
Nucl.\ Phys.\ B {\bf 429}, 375 (1994)
[arXiv:hep-th/9403141].
}

\lref\FZZ{
V.~Fateev, A.~B.~Zamolodchikov and A.~B.~Zamolodchikov,
``Boundary Liouville field theory. I: Boundary state and boundary  two-point function,''
arXiv:hep-th/0001012.
}

\lref\PST{
B.~Ponsot, V.~Schomerus and J.~Teschner,
``Branes in the Euclidean AdS(3),''
JHEP {\bf 0202}, 016 (2002)
[arXiv:hep-th/0112198].
}

\lref\PT{
B.~Ponsot and J.~Teschner,
``Boundary Liouville field theory: Boundary three point function,''
Nucl.\ Phys.\ B {\bf 622}, 309 (2002)
[arXiv:hep-th/0110244].
}

\lref\TLbound{
J.~Teschner,
``Remarks on Liouville theory with boundary,''
arXiv:hep-th/0009138.
}

\lref\TLtwo{
J.~Teschner,
``A lecture on the Liouville vertex operators,''
arXiv:hep-th/0303150.
}

\lref\TB{
J.~Teschner,
``Liouville theory revisited,''
Class.\ Quant.\ Grav.\  {\bf 18}, R153 (2001)
[arXiv:hep-th/0104158].
}

\lref\ZZ{
A.~B.~Zamolodchikov and A.~B.~Zamolodchikov,
``Structure constants and conformal bootstrap in Liouville field theory,''
Nucl.\ Phys.\ B {\bf 477}, 577 (1996)
[arXiv:hep-th/9506136].
}

\lref\TLconst{
J. Teschner,
``A lecture on the Liouville vertex operators,''
Preprint [arXiv:hep-th/0303150]
}

\lref\GaberdielXM{
M.~R.~Gaberdiel, A.~Recknagel and G.~M.~Watts,
``The conformal boundary states for SU(2) at level 1,''
Nucl.\ Phys.\ B {\bf 626}, 344 (2002)
[arXiv:hep-th/0108102].
}

\lref\tendim{
J.~A.~Harvey, S.~Kachru, G.~W.~Moore and E.~Silverstein,
``Tension is dimension,''
JHEP {\bf 0003}, 001 (2000)
[arXiv:hep-th/9909072].
}

\lref\ZZB{
A.~B.~Zamolodchikov and A.~B.~Zamolodchikov,
``Liouville field theory on a pseudosphere,''
arXiv:hep-th/0101152.
}

\lref\LLM{
N.~Lambert, H.~Liu and J.~Maldacena, ``Closed strings from
decaying D-branes,'' arXiv:hep-th/0303139.
}

\lref\GutperleXF{
M.~Gutperle and A.~Strominger,
``Timelike boundary Liouville theory,''
arXiv:hep-th/0301038.
}

\lref\HarveyNA{
J.~A.~Harvey, D.~Kutasov and E.~J.~Martinec,
``On the relevance of tachyons,''
arXiv:hep-th/0003101.
}

\lref\GinspargIS{
P.~Ginsparg and G.~W.~Moore,
``Lectures On 2-D Gravity And 2-D String Theory,''
arXiv:hep-th/9304011.
}

\lref\DasKA{
S.~R.~Das and A.~Jevicki,
``String Field Theory And Physical Interpretation Of D = 1 Strings,''
Mod.\ Phys.\ Lett.\ A {\bf 5}, 1639 (1990).
}

\lref\JevickiQN{
A.~Jevicki,
``Development in 2-d string theory,''
arXiv:hep-th/9309115.
}

\lref\CallanUB{
C.~G.~Callan, I.~R.~Klebanov, A.~W.~Ludwig and J.~M.~Maldacena,
``Exact solution of a boundary conformal field theory,''
Nucl.\ Phys.\ B {\bf 422}, 417 (1994)
[arXiv:hep-th/9402113].  The Nuclear Physics version
contains an additional note performing the Cardy calculation.
}

\lref\PolchinskiMY{
J.~Polchinski and L.~Thorlacius,
``Free Fermion Representation Of A Boundary Conformal Field Theory,''
Phys.\ Rev.\ D {\bf 50}, 622 (1994)
[arXiv:hep-th/9404008].
}

\lref\GinspargIS{
P.~Ginsparg and G.~W.~Moore,
``Lectures On 2-D Gravity And 2-D String Theory,''
arXiv:hep-th/9304011.
}

\lref\emil{
E.~J.~Martinec,
``The annular report on non-critical string theory,''
arXiv:hep-th/0305148.
}

\lref\teleology{ J.~Teschner, ``On boundary perturbations in
Liouville theory and brane dynamics in noncritical string theories,''
arXiv:hep-th/0308140.
}


\lref\BerensteinXK{
D.~Berenstein, R.~Corrado, W.~Fischler, S.~Paban and M.~Rozali,
``Virtual D-Branes,''
Phys.\ Lett.\ B {\bf 384}, 93 (1996)
[arXiv:hep-th/9605168].
}

\lref\FischlerJA{
W.~Fischler, S.~Paban and M.~Rozali,
``Collective Coordinates for D-branes,''
Phys.\ Lett.\ B {\bf 381}, 62 (1996)
[arXiv:hep-th/9604014].
}

\lref\PO{
V.~Periwal and O.~Tafjord,
``D-brane recoil,''
Phys.\ Rev.\ D {\bf 54}, 3690 (1996), arXiv:hep-th/9603156.
}

\lref\kogan{
I.~I.~Kogan and N.~E.~Mavromatos,
``World-Sheet Logarithmic Operators and Target Space Symmetries in String
Theory,''
Phys.\ Lett.\ B {\bf 375}, 111 (1996)
[hep-th/9512210];
I.~I.~Kogan, N.~E.~Mavromatos and J.~F.~Wheater,
``D-brane recoil and logarithmic operators,''
Phys.\ Lett.\ B {\bf 387}, 483 (1996)
[hep-th/9606102];
I.~I.~Kogan and J.~F.~Wheater,
``ND tadpoles as new string states and quantum mechanical particle-wave
duality from world-sheet T-duality,''
Phys.\ Lett.\ B {\bf 403}, 31 (1997)
[hep-th/9703141].
}

\lref\PolchinskiFQ{
J.~Polchinski,
``Combinatorics Of Boundaries In String Theory,''
Phys.\ Rev.\ D {\bf 50}, 6041 (1994)
[arXiv:hep-th/9407031].
}

\lref\PolyakovAF{
A.~M.~Polyakov,
``Gauge fields and space-time,''
Int.\ J.\ Mod.\ Phys.\ A {\bf 17S1}, 119 (2002)
[arXiv:hep-th/0110196].
}

\lref\poljoerg{J. Teschner, unpublished.}
\lref\kms{
I.~Klebanov, J.~Maldacena, N.~Seiberg,
hep-th/0305159.}
\lref\freeferm{
E.~Brezin, C.~Itzykson, G.~Parisi and J.~B.~Zuber,
``Planar Diagrams,''
Commun.\ Math.\ Phys.\  {\bf 59}, 35 (1978).
}

\lref\GarousiTR{
M.~R.~Garousi,
``Tachyon couplings on non-BPS D-branes and Dirac-Born-Infeld action,''
Nucl.\ Phys.\ B {\bf 584}, 284 (2000)
[arXiv:hep-th/0003122].
}

\lref\SenMD{
A.~Sen,
``Supersymmetric world-volume action for non-BPS D-branes,''
JHEP {\bf 9910}, 008 (1999)
[arXiv:hep-th/9909062].
}

\lref\MoellerVX{
N.~Moeller and B.~Zwiebach,
``Dynamics with infinitely many time derivatives and rolling tachyons,''
JHEP {\bf 0210}, 034 (2002)
[arXiv:hep-th/0207107].
}

\lref\SenNU{ A.~Sen, ``Rolling tachyon,'' JHEP {\bf 0204}, 048
(2002) [arXiv:hep-th/0203211].
}

\lref\SenIN{ A.~Sen, ``Tachyon matter,'' JHEP {\bf 0207}, 065
(2002) [arXiv:hep-th/0203265].
}

\lref\SenTM{ A.~Sen, ``Dirac-Born-Infeld Action on the Tachyon
Kink and Vortex,'' [arXiv:hep-th/0303057].
}

\lref\SenQA{ A.~Sen, ``Time and tachyon,'' [arXiv:hep-th/0209122].
}

\lref\SenAN{ A.~Sen, ``Field theory of tachyon matter,'' Mod.\
Phys.\ Lett.\ A {\bf 17}, 1797 (2002) [arXiv:hep-th/0204143].
}

\lref\SenVV{ A.~Sen, ``Time evolution in open string theory,''
JHEP {\bf 0210}, 003 (2002) [arXiv:hep-th/0207105].
}

\lref\ChenFP{ B.~Chen, M.~Li and F.~L.~Lin, ``Gravitational
radiation of rolling tachyon,'' JHEP {\bf 0211}, 050 (2002)
[arXiv:hep-th/0209222].
}

\lref\SenMS{
P.~Mukhopadhyay and A.~Sen,
``Decay of unstable D-branes with electric field,''
JHEP {\bf 0211}, 047 (2002)
[arXiv:hep-th/0208142].
}

\lref\Rey{S.~J.~Rey and S.~Sugimoto,
``Rolling Tachyon with Electric and Magnetic Fields -- T-duality approach,''
[arXiv:hep-th/0301049].
}

\lref\GutperleAI{ M.~Gutperle and A.~Strominger, ``Spacelike
branes,'' JHEP {\bf 0204}, 018 (2002) [arXiv:hep-th/0202210].
}

\lref\stromingeropen{ A.~Strominger, ``Open string creation by
S-branes,'' [arXiv:hep-th/0209090].
}

\lref\tbl{M.~Gutperle and A.~Strominger,
``Timelike Boundary Liouville Theory,''
[arXiv:hep-th/0301038].
}

\lref\Larsen{F.~Larsen, A.~Naqvi and S.~Terashima,
``Rolling tachyons and decaying branes,''
[arXiv:hep-th/0212248].
}

\lref\MaloneyCK{
A.~Maloney, A.~Strominger and X.~Yin,
``S-brane thermodynamics,''
[arXiv:hep-th/0302146].
}

\lref\BuchelTJ{
A.~Buchel, P.~Langfelder and J.~Walcher,
``Does the tachyon matter?,''
Annals Phys.\  {\bf 302}, 78 (2002)
[arXiv:hep-th/0207235];
A.~Buchel and J.~Walcher,
``The tachyon does matter,''
arXiv:hep-th/0212150.
}

\lref\LeblondDB{ F.~Leblond and A.~W.~Peet,
``SD-brane gravity fields and rolling tachyons,''
[arXiv:hep-th/0303035].
}

\lref\OkudaYD{ T.~Okuda and S.~Sugimoto, ``Coupling of rolling
tachyon to closed strings,'' Nucl.\ Phys.\ B {\bf 647}, 101 (2002)
[arXiv:hep-th/0208196].
}

\lref\Aref{
I.~Y.~Aref'eva, L.~V.~Joukovskaya and A.~S.~Koshelev,
``Time evolution in superstring field theory on non-BPS brane.
I: Rolling  tachyon and energy-momentum
conservation,''
arXiv:hep-th/0301137.
}

\lref\IshidaCJ{
A.~Ishida and S.~Uehara,
``Rolling down to D-brane and tachyon matter,''
JHEP {\bf 0302}, 050 (2003)
[arXiv:hep-th/0301179].
}

\lref\KlusonAV{
J.~Kluson,
``Time dependent solution in open Bosonic string field theory,''
arXiv:hep-th/0208028;
``Exact solutions in open Bosonic string field theory and
marginal  deformation in CFT,''
[arXiv:hep-th/0209255].
}

\lref\MinahanIF{
J.~A.~Minahan,
``Rolling the tachyon in super BSFT,''
JHEP {\bf 0207}, 030 (2002)
[arXiv:hep-th/0205098].
}

\lref\SugimotoFP{
S.~Sugimoto and S.~Terashima,
``Tachyon matter in boundary string field theory,''
JHEP {\bf 0207}, 025 (2002)
[arXiv:hep-th/0205085].
}

\lref\ReyXS{
S.~J.~Rey and S.~Sugimoto,
``Rolling tachyon with electric and magnetic fields: T-duality approach,''
arXiv:hep-th/0301049.
}

\lref\LambertHK{
N.~D.~Lambert and I.~Sachs,
``Tachyon dynamics and the effective action approximation,''
Phys.\ Rev.\ D {\bf 67}, 026005 (2003)
[arXiv:hep-th/0208217].
}

\lref\HughesBW{
J.~Hughes, J.~Liu and J.~Polchinski,
``Virasoro-Shapiro From Wilson,''
Nucl.\ Phys.\ B {\bf 316}, 15 (1989).
}

\lref\condensingrefs{
For a review see
A.~Sen,
``Non-BPS states and branes in string theory,''
arXiv:hep-th/9904207.
}

\lref\rollingrefs{ \eg\ A.~Sen, ``Rolling tachyon,'' JHEP {\bf 0204}, 048
(2002) [arXiv:hep-th/0203211];
``Tachyon matter,'' JHEP {\bf 0207}, 065
(2002) [arXiv:hep-th/0203265];
``Time evolution in open string theory,''
JHEP {\bf 0210}, 003 (2002) [arXiv:hep-th/0207105];
M.~R.~Garousi,
``Tachyon couplings on non-BPS D-branes and Dirac-Born-Infeld action,''
Nucl.\ Phys.\ B {\bf 584}, 284 (2000)
[arXiv:hep-th/0003122];
M.~Gutperle and A.~Strominger, ``Spacelike
branes,'' JHEP {\bf 0204}, 018 (2002) [arXiv:hep-th/0202210];
A.~Strominger, ``Open string creation by
S-branes,'' [arXiv:hep-th/0209090];
M.~Gutperle and A.~Strominger,
``Timelike Boundary Liouville Theory,''
[arXiv:hep-th/0301038];
F.~Larsen, A.~Naqvi and S.~Terashima,
``Rolling tachyons and decaying branes,''
[arXiv:hep-th/0212248];
A.~Maloney, A.~Strominger and X.~Yin,
``S-brane thermodynamics,''
[arXiv:hep-th/0302146];
A.~Buchel and J.~Walcher,
``The tachyon does matter,''
arXiv:hep-th/0212150;
T.~Okuda and S.~Sugimoto, ``Coupling of rolling
tachyon to closed strings,'' Nucl.\ Phys.\ B {\bf 647}, 101 (2002)
[arXiv:hep-th/0208196];
A.~Ishida and S.~Uehara,
``Rolling down to D-brane and tachyon matter,''
JHEP {\bf 0302}, 050 (2003)
[arXiv:hep-th/0301179];
S.~Sugimoto and S.~Terashima,
``Tachyon matter in boundary string field theory,''
JHEP {\bf 0207}, 025 (2002)
[arXiv:hep-th/0205085];
B.~Chen, M.~Li and F.~L.~Lin,
``Gravitational radiation of rolling tachyon,''
JHEP {\bf 0211}, 050 (2002)
[arXiv:hep-th/0209222];
N.~D.~Lambert and I.~Sachs,
``Tachyon dynamics and the effective action approximation,''
Phys.\ Rev.\ D {\bf 67}, 026005 (2003)
[arXiv:hep-th/0208217].
}

\lref\recentrolling{
Y.~Demasure and R.~Janik, hep-th/0305191.
}

\lref\neil{
N.~Constable and F.~Larsen, hep-th/0305177.
}

\lref\igor{
I.~R.~Klebanov,
``String theory in two-dimensions,''
arXiv:hep-th/9108019.
}

\lref\SenMH{
A.~Sen,
``Descent relations among bosonic D-branes,''
Int.\ J.\ Mod.\ Phys.\ A {\bf 14}, 4061 (1999)
[arXiv:hep-th/9902105].
}

\lref\BanksQS{
T.~Banks and E.~J.~Martinec,
``The Renormalization Group And String Field Theory,''
Nucl.\ Phys.\ B {\bf 294}, 733 (1987).
}

\lref\ZamolodchikovGT{
A.~B.~Zamolodchikov,
``'Irreversibility' Of The Flux Of The Renormalization Group In A 2-D Field Theory,''
JETP Lett.\  {\bf 43}, 730 (1986)
[Pisma Zh.\ Eksp.\ Teor.\ Fiz.\  {\bf 43}, 565 (1986)].
}


\Title{\vbox{\baselineskip3pt\hbox{hep-th/0305194}
\hbox{PUTP-2087}}}
{\vbox{ \centerline{Classical and Quantum D-branes }}}
\vskip -15pt
{\vbox{ \centerline{\titlefont in 2D String Theory}}}

\bigskip

\bigskip
\bigskip
\centerline{John McGreevy,${}^1$
 Joerg Teschner${}^2$ and Herman Verlinde${}^1$}
\bigskip
\bigskip
\bigskip
\centerline{${}^1${\it Department of Physics, Princeton University,
Princeton, NJ 08544}}
\bigskip
\centerline{${}^2${\it Institut f\"ur theoretische Physik, Freie Universit\"at
Berlin, 14195 Berlin}}
\bigskip
\bigskip
\bigskip
\noindent
We investigate two classes of D-branes in 2-d string theory,
corresponding to extended and localized branes, respectively.
We compute the string emission during tachyon condensation and reinterpret
the results within the $c=1$ matrix model. As in hep-th/0304224, we find
that the extended branes describe classical eigenvalue trajectories, while,
as found in hep-th/0305159, the localized branes correspond to the
quantum field that creates and destroys eigenvalues.
The matrix model relation between the classical probe and the local collective
field precisely matches with the descent relation between
the boundary states of D-strings and D-particles.

\bigskip
\Date{May, 2003}

\newsec{Introduction}

D-branes are fascinating objects. With the benefit of hindsight,
it appears
surprising that it took such a long time before the full extent of
their role in string theory was appreciated. Undoubtedly, part of
their true meaning was obscured by the fact that D-branes start
their perturbative life as unnaturally rigid looking objects, with
fixed positions or following a fixed classical trajectory.
They fully come to life, however, after one realizes that the
open/closed string dynamics on their world-volume promotes all
their properties, and in particular their positions, into
dynamical, quantum mechanical degrees of freedom. A quantum
D-brane not only affects but also reacts to its environment.
Including such recoil effects is particularly important in the
study of time-dependent phenomena, such as the decay of unstable
D-branes into closed strings \rollingrefs.

A useful arena for studying D-brane dynamics in a fully quantum
mechanical context is 2-dimensional bosonic string theory, which
is believed to have an exactly soluble dual description in terms
of matrix quantum mechanics. In \McGreevyKB\ it was proposed that
the hermitian $c=1$ matrix in fact represents the open string
tachyon mode of a dense gas of unstable D-particles. In light of
this conjecture, it is natural to suspect that the properties and
interactions of D-branes in 2-d string theory must have a simple
and natural representation in terms of the matrix eigenvalues.
With this dictionary in hand, one can then proceed to learn new
lessons about the quantum properties of D-branes, some of which
hopefully generalize to arbitrary string backgrounds.

In this paper we will investigate two classes of D-branes in 2-d
string theory, corresponding to extended and localized branes,
respectively. We describe their open string spectrum and the
closed string one-point functions on the disk that quantify the
perturbative string emission during tachyon condensation. We then
reinterpret the results within the $c=1$ matrix model. We clarify
and extend of the result of \McGreevyKB\ that the extended branes
describe the response to classical probe-eigenvalue trajectories.
The localized brane, on the other hand, has the matrix model
interpretation as the quantum field that creates and destroys
eigenvalues. This interesting result, which was recently obtained
in \kms, suggests that the localized brane is closely related to
the operator that creates the quantum version of the extended
brane.

\subsec{Classical versus Quantum D-branes}

For our later discussion, it will be useful to clarify what we
mean by classical and quantum D-branes. Abstractly, a D-brane is
specified by a conformally invariant boundary state $|B_X\rangle$,
which acts as a linear source for all closed string fields
$\phi_i$: \eqn\boundary{ S_{int}(\phi,X) =\sum_i \, \phi_i \,
\langle V_i | B_X\rangle } Here $X$ denotes some set of parameters
that specify the boundary CFT.

This characterization of a D-brane by a boundary state requires
that the profiles of the open-string modes on the brane satisfy
their classical equations of motion: the D-brane is still
classical. Clearly, however, any object responds to its
environment, and consistency of string perturbation theory
dictates that all brane degrees of freedom must be quantized
\PolchinskiFQ.
The quantization of the collective coordinates
of D-branes
was further elaborated in
\refs{\FischlerJA,\BerensteinXK,\PO,\kogan},
and, of course,
is crucial for the existence of
quantum field theories on the worldvolumes of branes.

For definiteness, let us specialize to the case where the D-brane
is a D-particle, and $X$ denotes its position. Its possible
boundary states correspond to all possible classical trajectories
$X(t)$. To quantize the particle, however, it would be helpful if
we could consider arbitrary time dependent trajectories $X(t)$, so
that we can write a path-integral expression for its wavefunction
as \eqn\bigpsi{\Psi(X) = \int\limits^{X}\!\! DX(t)\; e^{{i\over
\hbar} \bigl( S(X)+ S_{int}(\phi,X)\bigr)}\, ,} where $S(X)$ is
the D-particle worldline action in the string background $\phi_i
=0$. The interaction term $S_{int}$ must be a generalization of
\boundary\ to general off-shell trajectories, so that we can write
\eqn\sint{S_{int} = \int \!\! dt \, L_{int}(\phi,X(t)) \, .} In
this paper, motivated by the matrix model formulation of 2-d
string theory, we will adopt the following general prescription
for obtaining $L_{int}$, which works for general static target
space-times. Let $|B_{X^i}\rangle$ denote the boundary state in
the space-like CFT describing the D-particle at location $X^i$,
and let $|B^D_{X^0=t}\rangle$ denote the time-like Dirichlet
boundary state at $X^0=t$. Then we define
 \eqn\lint{L_{int}(\phi,X(t)) = \sum_i \, \phi_i \,
\langle V_i | B_{X(t)} \rangle}
 \eqn\boundaryt{ | B_{X(t)}\rangle = |B_{X^i(t)}\rangle \otimes
|B^D_{X^0=t}\rangle} Here in the first factor on the right-hand
side, $X_i(t)$ is just a fixed  position and not the full
time-dependent trajectory; the state $|B_{X^i(t)}\rangle$ defines
a consistent boundary CFT. So, via this proposal we can consider
arbitrary D-brane trajectories $X(t)$, while keeping the power of
exact worldsheet conformal invariance.

The D-particle action and wave-function \bigpsi\ both explicitly
depend on the closed string background parameterized by the
$\phi_i$. Now suppose we promote the $\phi_i$ to quantum fields,
and in addition introduce a multi-particle Hilbert space for the
D-particles. In this way, we can try to promote the D-particle
wave-function $\Psi(X)$ to a second quantized field operator,
which we would like to identify with the ``gravitationally
dressed'' version of the local quantum field that creates and
destroys D-particles. This we would like to call the quantum
D-brane.

\bigskip




\subsec{Organization}

The paper is organized as follows.  In the next section,
we study the Liouville CFT and its boundary states in the limit
($c \to 25$) in which it participates in two-dimensional bosonic string theory.
In \S3, we embed this discussion in 2D string theory
by studying tensor products of Liouville boundary states
with various possible boundary states of the $X^0$ CFT.
In this context, we apply the procedure described in \S1.1
to study D-branes on arbitrary trajectories.
Along the way, we perform a
(logically independent)
Cardy analysis of the
spectral density of open-strings associated with the
bouncing boundary state, which
corroborates our earlier discussion.
In \S4 and \S5 we turn to the matrix model description of these
processes, using classical and quantum descriptions
of probe eigenvalues.
We conclude in \S6.

\bigskip

While we were typing this sentence,
\emil\ appeared, which in addition to many
other interesting results, identifies in detail
a matrix model counterpart for the
ZZ states with $m,n > 1$.


\newsec{Boundary Liouville CFT}

As usual we will define the 2d string theory as the tensor product of
the CFT of a free boson with the $c=25$ Liouville theory.
In order to define the latter we shall take the limit $c\downarrow 25$ of the
$c>25$ Liouville theory constructed in \TB\TLtwo.

In the semiclassical limit $c\ra\infty$ one may describe Liouville
theory in terms of the action
\eqn\cL{
S_{\rm cl}\;=\;
\int d^2z \biggl(\frac{1}{\pi}|\pa_z\varphi|^2+\mu e^{2b\varphi}\biggl)\;.}
The basic fields of the theory are the primary fields
\eqn\vaa{\qquad \qquad
V_{\al}(z,\bz)\simeq e^{2\al\varphi(z,\bz)}, \qquad \qquad {b\ra \infty}},
which have
conformal dimensions $\Delta_{\al}=\al(Q-\al)$. It will be important for
us to remember that these fields satisfy the reflection property \ZZ
\eqn\refl{
V_{\frac{Q}{2}+iP}(z,\bz)=R(P)V_{\frac{Q}{2}-iP}(z,\bz),
} where
\eqn\refltwo{
R(P)\;=\; -(\pi\mu\gamma(b^2))
^{-\frac{2iP}{b}}\frac{\Ga(1+2ibP)\Ga(1+2ib^{-1}P)}
{\Ga(1-2ibP)\Ga(1-2ib^{-1}P)}\;.
}
with
\eqn\gadef{
\gamma(b^2) \equiv \frac{\Ga(b^2)}{\Ga(1-b^2)}\;.
}
There are subtle quantum modifications of the action \cL\
for finite values of $c$. In order to describe the behavior
of Liouville theory in the weak-coupling asymptotics $\varphi\ra -\infty$
one may use the following action (as explained in Part II of \TB)
\eqn\qL{
S_{\rm q}\;=\;\int d^2z \biggl(\frac{1}{\pi}|\pa_z\varphi|^2+\mu e^{2b\varphi}+\tilde{\mu}
e^{2b^{-1}\varphi}\biggl)\;,
}
where the coupling constant $b$ is related to $c$ via
$c=1+6Q^2$, $Q=b+b^{-1}$, and $\tilde{\mu}(\mu,b)$ is given by
\eqn\xxx{
\pi\gamma(b^{-2})\tilde{\mu}=(\pi \gamma(b^2)\mu)^{b^{-2}}.
}
In order to get a useful perspective
on the quantum corrections appearing in \qL\ let us note that
\eqn\qLprime{
S_{\rm q}\;=\;
\int d^2z \biggl(\frac{1}{\pi}|\pa_z\varphi|^2+\mu V_b(z,\bz)\biggl)\;.
}
The correspondence with \qL\ follows from the observations that for $\varphi \ra -\infty$
\eqn\xxx{
\qquad V_{\al}\;{\simeq}
e^{2\al\varphi}+R_{\al}
e^{2(Q-\al)\varphi}, \qquad
R_{\al}\equiv R\left(\fr{i}{2}(Q-2\al)\right).}
and that $R_b\equiv \tilde{\mu}/\mu$.

We are interested in the limit $c\ra 25$, corresponding to $b\ra 1$.
In order to get finite results for the basic quantities like the
reflection amplitude \refltwo\ or the three-point function
\DO\ZZ\ we need to keep the combination
\eqn\yyy{
\mu_{\rm ren}\equiv \mu\gamma(b^2)
}
finite in the limit $b\ra 1$. Concerning the action $S_{\rm q}$ let us
note that
\eqn\xxx{
R_b\sim -(\pi \mu_{\rm ren})^{\ep}+\CO(\ep^2), \quad \ep\equiv 1-b^2,
}
which implies
 \eqn\cosmas{
\eqalign{
\mu\big(e^{2b\varphi}+R(b)e^{2b^{-1}\varphi}\big)\;\;\sim
\;\;\mu\big(e^{-\ep\varphi}-e^{\ep(\varphi+\ln\pi\mu_{\rm ren})}\big)e^{2\varphi}\cr
\qquad \qquad \sim  \;\;-\mu_{\rm ren}(2\varphi+\log\pi\mu_{\rm ren})e^{2\varphi}.}}
In this way we have derived some old conjectures
concerning the form
of the Liouville-interaction for $c=25$ from the exact solution.

It should be emphasized, however, that \cosmas\ serves to
describe the c=25 Liouville theory {\it only}
in the asymptotics $\varphi\ra \infty$; it can not be expected to
be exact \TB.

\def\mub{\mu_{\rm B}}

\subsec{Boundary Liouville theory at $c>25$}

A first important class of conformally invariant boundary conditions for the
$c>25$ Liouville theory may be defined in the semiclassical limit
$c\ra\infty$ by the boundary action
\eqn\yyy{
S_{\rm bound}\;=\;\int_{\pa\Ga}\frac{d\tau}{2\pi} \;g^{\frac{1}{4}}\bigl(
QK+2\pi \mu_{\rm B}e^{b\varphi}\bigr),
}
where $\tau$ is a parameter for the boundary, and $K$ is its extrinsic
curvature in the background metric $g$.
The parameter $\mub$ which labels the boundary conditions
is called the boundary cosmological constant.

\def\De{\Delta}
\def\ga{\gamma}
\def\sst{{}}

The corresponding boundary states were constructed in \FZZ.
They can be represented as
\eqn\boundst{
|B_s\rangle \;=\;\int\limits_{-\infty}^{\infty}\frac{dP}{2\pi}\, e^{-2\pi iPs}
v(P) | \, P\rangle\!\rangle,
}
where the following definitions have been used:
$|\, P\rangle\!\rangle$ is the Ishibashi-state built from the Virasoro
representation with highest weight $\Delta_P=\frac{1}{4}(Q^2+4P^2)$,
the function $v(P)$ is given as
\eqn\xxx{
v(P)=i\bigl(\pi\mu\ga(b^2)\bigr)^{\frac{iP}{b}}
\frac{\Ga(1-2ibP)\Ga(1-2iP/b)}{P}\;,
}
and the parameter $s$ used in \boundst\ is related to the
boundary cosmological constant $\mub$ via
\eqn\smub{
\cosh\pi b s=\frac{\mub}{\sqrt{\mu}}\sqrt{\sin\pi b^2}\;.
}
Using the parameter $s$ instead of
$\mub$ turns out to be rather natural for the
description of boundary Liouville theory.

In order to construct open string theories involving a Liouville
direction, one needs to consider Liouville theory on the strip
$[0,\pi]$ with boundary conditions labelled by
parameters $s_2$ and $s_1$. The spectrum $\CH_{s_2s_1}^{\rm B}$
of Liouville theory
on the strip can be deduced from the bulk spectrum
via world-sheet duality \TLbound. It always contains a
continuous part given by $\int_{0}^{\infty}dP~ \CV_{P,c}$, where
$\CV_{P,c}$ is the highest weight representation of the Virasoro algebra
with weight $\De_P$ and central charge $c$.
In addition to the bulk primary fields $V_{\al}(z,\bz)$ one may now also
consider the boundary fields $\Psi^{s_2s_1}_{\al}(\tau)$
which create the states $|\al;s_2,s_1\rangle_{\rm\sst B}$
in $\CH_{s_2s_1}^{\rm\sst B}$  from the
vacuum. The boundary fields are fully characterized
by their covariance w.r.t. conformal transformations together
with their three point function on the disk \PT.
They satisfy a reflection property analogous to \refl :
\eqn\brefl{
\Psi^{s_2s_1}_{\frac{Q}{2}+iP}(\tau)=R\bigl(P|s_2,s_1\bigr)
\Psi^{s_2s_1}_{\frac{Q}{2}-iP}(\tau)\;,
}
where the boundary reflection amplitude $R\bigl(P|s_2,s_1\bigr)$
is given by the expression \FZZ
\eqn\xxx{\eqalign{
{}  R\bigl(P|s_2,s_1\bigr)\;=\;&
\bigl(\pi \mu \ga(b^2) b^{2-2b^2}\bigr)^{-\frac{iP}{b}}\;
\frac{\Ga_b(+2iP)}{\Ga_b(-2iP)}\times \cr
& \times\frac{
 S_b\bigl(\fr{Q}{2}-i(P+s_1+s_2)\bigr)
 S_b\bigl(\fr{Q}{2}-i(P+s_1-s_2)\bigr)}
{S_b\bigl(\fr{Q}{2}+i(P-s_1-s_2)\bigr)
 S_b\bigl(\fr{Q}{2}+i(P-s_1+s_1)\bigr)}}}
Integral representations for the special functions $G_b$ and
$S_b$ can be found \eg\ in \FZZ. For our purposes it will be enough
to note that $G_b(x)$ and
$S_b(x)$ are analytic in the strip $0<\Re(x)<Q$ and have a simple pole
at $x=0$.

\def\la{\lambda}
\def\mur{\mu_{\rm\sst ren}}
\def\murb{\mu_{\rm\sst B,ren}}

Another very interesting class of boundary states were found in \ZZB.
At present we only have a physical interpretation for the simplest of
these boundary states, which will be denoted $|B_{\rm ZZ}\rangle$.
It is given by an expression of the following form:
\eqn\boundstZZ{
|B_{\rm ZZ}\rangle \;=\;\int\limits_{-\infty}^{\infty}\frac{dP}{2\pi}\,
U_{\rm ZZ}(P) | \, P\rangle\!\rangle,
}
where the one-point function
$U_{\rm ZZ}(P)\equiv \langle P|B_{\rm ZZ}\rangle$ is
\eqn\oneptZZ{
U_{\rm ZZ}(P)=
\frac{\big(\pi \mu \ga(b^2)\big)^{(2iP-Q)/2b}\Ga(bQ)\Ga(Q/b)Q}
{2iP\;\Ga(2ibP)\Ga(2iP/b)}\;.
}
The boundary conditions described by \boundstZZ\ have the remarkable
property that the spectrum of Liouville theory on the strip with
boundary conditions corresponding to $|B_{\rm ZZ}\rangle$ on both sides
contains the Virasoro representation of the identity {\it only} \ZZB.

\subsec{Boundary Liouville theory at $c=25$}

Our aim in this subsection is to discuss the
limit $b\ra 1$ of boundary Liouville theory. Taking this limit is
unproblematic in the case of $|B_{\rm ZZ}\rangle$.
The one-point function for the ZZ state becomes
\eqn\ZZbtoone{
U_{ZZ}(P) = {2 \over i\pi }e^{i \delta(2 P)} \sinh 2 \pi P,
}
where the phase
\eqn\legpole{
e^{i\delta(P)} = (\pi \mur)^{-iP/ 2} {\Gamma(1+iP)
\over \Gamma(1-iP)}}
is known as the legpole factor.

However, taking $b \to 1$ requires a bit of
care in the case of the states $|B_s\rangle$.
Let us first of all note that in the
limit $b\ra 1$ the equation \smub\ becomes
\eqn\xxx{
\cosh \pi s\;
\simeq\;\mub\sqrt{\frac{\ep}{\mur}}\sqrt{\ep}\;\equiv
\frac{\murb}{\sqrt{\mur}},
}
introducing a renormalized boundary cosmological constant $\murb$ that
corresponds to finite values of $s$. A useful perspective
on the origin of the renormalization of $\mub$ can be gained from the
following considerations.
The semiclassical field $e^{b\varphi(\tau)}$ that appears in the
boundary action gets replaced by the
primary boundary field $\Psi^{s,s}_{b}(\tau)$ in the
quantum theory. Indeed, taking into account
the reflection relation \brefl\ we may observe that
$\Psi^{s,s}_{b}(\tau)$ is the {\it unique} primary boundary
field with conformal dimension 1.
We may then notice that for $\varphi \to -\infty$
\eqn\bcosmas{
\Psi^{s,s}_{b}(\tau)
\;\simeq \;e^{b\varphi}+R_{b,s}~e^{b^{-1}\varphi},\qquad \ \
R_{\al,s}\equiv R\left(\fr{i}{2}(Q-2\al);s,s\right)\;.
}
Upon taking the limit $b\ra 1$ we now have
\eqn\xxx{
R(b|s,s)\sim -(\pi \mu_{\rm ren})^{\ep/2}+\CO(\ep^2).
}
The important minus sign in front comes from the factor
$\Ga_b(2iP)/\Ga_b(-2iP)$ in the boundary reflection amplitude.
Inserting this into
\bcosmas\ yields
 \eqn\bcosren{
\eqalign{
\mub\big(e^{b\varphi}+R(b|s,s)e^{b^{-1}\varphi}\big)\;\;\sim &
\;\;\mub\big(e^{-\frac{1}{2}\ep\varphi}-
e^{\frac{1}{2}\ep(\varphi+\log\pi\mu_{\rm ren})}\big)e^{\varphi}\cr
\sim & \;\;-\murb(\varphi+\fr{1}{2}\log\pi\mur)e^{\varphi}\;.
}}

\noindent{\bf Remarks}

\item{1.} More formally one may deduce the need for renormalization
of $\mub$ from the fact that $\Psi_b^{s,s}$ vanishes as $1-b^2$
for $b\ra 1$, as follows from the results of \PT. We may then
define $\murb$ such that
\eqn\renorm{
\lim_{b\ra 1}\mub \Psi_{b}^{s,s}(x)=
\murb\Psi_1'{}^{s,s}(x),\qquad
\Psi_1'{}^{s,s}(x)\equiv
\big[\pa_{\al}^{}\Psi_{\al}^{s,s}(x)\big]_{\al=1}^{}\;.
}
\item{2.} We would like to emphasize that the existence of a
continuous spectrum in $\CH_{s_2s_1}^{\rm\sst B}$ is unaffected
by the limit $b\ra 1$. It follows rather robustly from the pole
at $P=0$ in the boundary state \boundst. The corresponding
divergence of the annulus amplitude is related by
world-sheet duality to the usual volume divergence that
signals the presence of continuous spectrum in the
open string channel; see \eg\ \PST\
for a more detailed discussion in a very similar context.
The results of \TLbound\ furthermore imply absence of bound states in
$\CH_{s_2s_1}^{\rm B}$ as long as $s_i$, $i=1,2$, correspond
to positive values of $\murb$, \ie\
\eqn\spec{
\CH_{s_2s_1}^{\rm B}=
\int_{0}^{\infty}dP\; \CV_{P,25}\;,
}
where $\CV_{P,c}$ denotes the Virasoro highest-weight
representation with central charge $c=1+6Q^2$
and highest weight $\De_P=\frac{1}{4}Q^2+P^2$.

\item{3.} Just as for \cosmas, we should emphasize that \bcosren\ is
{\it not} expected to be accurate at finite values
of $\varphi$; we only expect it to be useful in the asymptotics
$\varphi\ra -\infty$.

\item{4.} Nevertheless we may infer the main qualitative features
of the boundary interaction as follows. If the world-sheet is represented
as the upper half-plane we may represent the boundary condition
as
\eqn\bcond{
i(\pa-\bar{\pa})\varphi(x)=2\pi b \murb \Psi_1'{}^{s,s}(x),
}
where $\Psi_1'{}^{s,s}(x)$ was defined in \renorm\ and
$\varphi(z,\bz)\equiv\frac{1}{2}\big[\pa_{\al}V_{\al}(z,\bz)\big]_{\al=0}^{}$.
Equation \bcosren\ implies that the boundary conditions tend
to Neumann-type boundary conditions in the limit $\varphi\ra-\infty$.
On the other hand let us note that the reflection property
\brefl\ implies that the boundary potential still has the property that
it is fully reflecting\foot{Strictly speaking, we should tune the
bulk potential to zero to make this assertion.
By using the results of \PT\ it is possible
to show that taking $\mu\ra 0$ preserves the
reflection relation \brefl.}.

\newsec{D-branes in 2d string theory}

\subsec{Extended D-branes}

A certain class of D-branes in the two-dimensional string theory
may be described by boundary potentials of the form
\eqn\genpots{
T_{\rm roll,\nu}(X^0,\varphi)=e^{\nu X^0}e^{2(1-\nu)\varphi},\qquad
T_{\rm bounce,\nu}(X^0,\varphi)=\cosh(\nu X^0)e^{2(1-\nu)\varphi}\;.
}
Our aim will be to gain insight into the space-time interpretation of
these branes. Of course one can only hope for an
exact solutions of the
corresponding boundary CFT's for
particular choices of the parameter
$\nu$, like $\nu=0$ or $\nu=1$. Linear combinations of the resulting
potentials will correspond to boundary states of the form
\eqn\xxx{
|B\rangle\;=\; |B_{X^0}\rangle\ot|B_s\rangle\;,
}
where $|B_{X^0}\rangle$ is some boundary state in the
free boson CFT that corresponds to one of the following boundary potentials
\eqn\xxx{
T_{\rm static}(X^0)=0,\qquad T_{\rm bounce}(X^0)=\lambda \cosh X^0, \qquad
T_{\rm roll}(X^0)=\lambda e^{X^0}\;.
}
These boundary states of the $X^0$ CFT have been
the subject of active investigation
\rollingrefs.

The space-time interpretation of the corresponding branes is of course
simplest in the case of $T_{\rm static}(X^0)$,
where the time direction is represented simply by a free boson
with Neumann type boundary conditions. It has spectrum
$\int_{-\infty}^{\infty}d\omega\; \CV_{\omega,1}$. Tensoring
with Liouville theory yields a continuous spectrum of open
strings parametrized by the real half-line.

Our previous discussion of boundary Liouville theory then
provides the space-time interpretation of these branes.
Let us recall that the
boundary conditions that correspond to the $|B_s\rangle$-boundary states
tend to the the Neumann boundary conditions for $\varphi\ra -\infty$.
In this limit we therefore find D-strings
stretched along the $\varphi$-direction. The reflection property \refl\
implies that the force that
acts upon the end points of the open strings ultimately pushes
all of them into the weak coupling region $\varphi\ra -\infty$.
It is natural to interpret this fact by saying that the
D-strings gradually disappear when we go into the strong
coupling region $\varphi\ra +\infty$: We need open strings with
very high energies in order to probe deeply into $\varphi\ra +\infty$.

To generalize slightly one may imagine taking the limit where
$\nu\ra 0$ in the potentials \genpots. Since the variation with
$X^0$ becomes very slow one would expect that this limit
corresponds to the adiabatic approximation. This may be supported by
observing that $X^0$ is related to the corresponding classical field
$X^0_{\rm cl}$ by a rescaling $X^0_{\rm cl}\equiv\sqrt{\hbar}X^0$.
Identifying  $\nu\equiv \sqrt{\hbar}$ we observe that sending
$\nu\ra 0$ should allow us to replace $\nu X^0$ by its classical value
$X^0_{\rm cl}$. We are thereby lead to expect that the adiabatic
approximation simply means making the boundary cosmological constant $\murb$
time-dependent. We will return to this situation in \S3.3.

The situation is somewhat more subtle in the cases where the boundary
potential involves
$T_{\rm bounce}(X^0)$ or $T_{\rm roll}(X^0)$. The work of
\CallanUB\GaberdielXM\
gives good control over the corresponding Euclidean theories,
but the analytic continuation of these results to the
corresponding timelike CFT's is subtle and does not seem to be
fully understood. In \S3.4 we will discuss
this analytic continuation for
the case of  $T_{\rm bounce}(X^0)$. For the time being let us consider the
case of $T_{\rm roll}(X^0)$, which seems much simpler in this respect.
In this case we have the one-point
function \LLM
\eqn\xxx{
\langle\omega|B_{\la}\rangle^{}_{\rm\sst roll}=\pi
\frac{ \lambda^{-i\omega} }{\sinh(\pi\omega)},
}
which displays a pole at $\omega=0$. Following our Remark 2 at the end
of the previous section we therefore expect that the spectrum contains
$\int_{0}^{\infty}d\omega\; \CV_{\omega,1}$. This would imply the existence
of a continuous spectrum of on-shell open strings. We will return to the
space-time interpretation of $T_{\rm roll}(X^0)$ in
\S3.3.

\subsec{Remarks on the tension of D-strings}

A reasonable concept of
``mass'' or ``tension'' for these branes is not obvious;
standard
discussions such as \tendim\ are not applicable here. Instead we
shall propose the following arguments.

After all, our D-strings are not homogeneous. In the weak coupling
region $\varphi \ra -\infty$ they are just like ordinary D-strings,
but if we go down to strong coupling $\varphi \ra \infty$ the
D-strings gradually disappear: Open strings will not have their
end-points in this region, as these end points would feel a strong
force pushing them into the weak coupling region.

We should therefore not expect the corresponding density of tension
to be homogeneous either. It should become the usual
constant tension per length of the D-string for $\varphi \ra -\infty$
and should vanish for $\varphi \ra \infty$. To have an estimate
for the total tension of the D-string we first of all need to
regularize the infinity from $\varphi \ra -\infty$ by introducing
some cut-off $\Lambda$. We may then try to estimate the tension by
replacing the boundary potential that makes the brane disappear
by a reflecting wall at $\varphi=\varphi_m$ with $\varphi_m(\mu_{B,ren})$
given by
\eqn\xxx{
\mu_{B,ren}\big(\varphi_m+\fr{1}{2}\log(\mu_{ren})
\big)e^{\varphi_m}=\sqrt{\mu_{ren}}.
}
This estimate may not be accurate quantitatively, but qualitatively
it seems clear that large values of $\mu_{B,ren}$ imply
large negative values of $\varphi_m$.

In order to support our proposal let us consider the overlap between
the boundary state $|B_s\rangle$ and a closed string wave-packet $\langle\Psi|$
that decays exponentially for $\varphi\ra-\infty$:
\eqn\xxx{
\langle\Psi|B_s\rangle=\int_{0}^{\infty}\frac{dP}{2\pi}\;\bar{\Psi}(P)
\;\langle P|B_s\rangle.
}
Exponential decay for $\varphi\ra-\infty$ implies that $\bar{\Psi}(P)$
is analytic in some strip around the real $P$-axis. The reflection
property \refl\ furthermore implies that
\eqn\xxx{
\bar{\Psi}(P)=\frac{1}{2}\big(\bar{\Psi}(P)+R(P)\bar{\Psi}(P)\big)
\sim ({\rm const.}) P \quad {\rm for}\;\, P\ra 0,
}
so that the pole of $\langle P|B_s\rangle$ gets cancelled. In the limit
$s\ra\infty$ corresponding to $\murb\ra\infty$ we therefore get
$\langle\Psi|B_s\rangle \ra 0$ from the factor $e^{-2\pi i sP}$ in
\boundst. This means that the coupling to
{\it all} closed string wave packets that decay fast in the
weak coupling region  $\varphi\ra-\infty$ goes to zero if we raise
the boundary cosmological constant to infinity.

We are thereby lead to the proposal that increasing $\murb$ (to infinity)
decreases the ``tension''. This would imply that the D-strings
are instable against processes which increase $\murb$. The rolling
tachyon discussed in \McGreevyKB\ is such a process as we will discuss
more explicitly in the following subsection.

\subsec{Rolling Tachyons: Worldsheet treatment}

In order to give an alternative description for the time
dependence of the background characterized by $T_{\rm
roll}(X^0)=\lambda e^{X^0}$ let us study a D-string with
time-dependent $\mu_{B,ren}\equiv z(t)=\la e^t$, considered as an
external source for closed strings. A natural ansatz for the
emission amplitude at a {\it fixed} time $t$ is given by the
overlap $\langle\omega|B_t\rangle$, where \eqn\Btbdstate{ |
B_t\rangle =| B^{\rm\sst D}_{t}\rangle^{}_{X^0}\ot|
B_{s(t)}\rangle,\qquad\quad \langle\omega|=
\big(\langle\omega|^{}_{X^0}\ot\langle P|\big)^{}_{P=\omega}\;. }
In \Btbdstate\ we have denoted the boundary state that realizes
Dirichlet boundary conditions for the time direction by $\langle
B^{\rm\sst D}_{t}|$ , \eqn\xxx{ \langle\omega| B^{\rm\sst
D}_{t}\rangle^{}_{X^0}=e^{-i\omega t}. } The total amplitude for
closed string emission is then given as \eqn\CAtimeint{
\CA(\omega)=\int\limits_{-\infty}^{\infty}dt\;\langle\omega|B_t\rangle
=\int\limits_{-\infty}^{\infty}  dt\,\, e^{i(\delta(\omega)-\omega
t)} { \cos \pi \omega \, s(t)\over \sinh  \pi \omega}\;, } where
$s(t)$ parameterizes the probe trajectory via \eqn\zpar{ z(t) =
\sqrt{\mu}\cosh \pi s(t) \;. } The rest of the calculation
proceeds as in \McGreevyKB: Let us change variables to $t=t(s)$,
defined by $\la e^{t(s)}=\sqrt{\mu}\cosh \pi s$. In doing this we
pick up a measure \eqn\xxx{
\rho(s)=\frac{dt}{ds}=\frac{\pi}{1+e^{2\pi
s}}-\frac{\pi}{1+e^{-2\pi s}}, } so that \eqn\owell{
\CA(\omega)=\int_{-\infty}^{\infty}  ds\,\rho(s)\,
e^{i(\delta(\omega)-\omega t(s))} { \cos \pi \omega \, s\over
\sinh  \pi \omega}\;. } We now comment on the form of this
integral.

The first thing we notice is that, as it stands, \owell\ is
infinite: the integrand at late and early times reduces to a
constant plus an oscillating piece, and the constant piece leads
to a linearly divergent contribution. The physical origin of this
divergence is that we are in effect sitting on top of a
resonance. We can regulate it by going a little bit off-shell,
replacing $e^{-i\omega t}$ by $e^{-i\tilde\omega t}$ with
$\tilde\omega$ slightly different from $\omega$. After this, it
should be possible to evaluate the integral by contour
deformation. We leave this task for the
future.

It is instructive to compare the integral \owell\ with the
emission amplitude due to the on-shell boundary CFT description of
the same tachyon condensation process. Consider the  special
boundary state $|B_{s={i\over 2}}\rangle$ tensored with the
rolling tachyon state. The corresponding on-shell production
amplitude \eqn\newroll{ \tilde{\CA}
(\omega)=\big[\langle\omega|B_{\lambda}\rangle^{}_{X^0} \langle
P|B_\frac{i}{2}\rangle\big]_{2P=\omega}^{}\;, } is equal to
\eqn\newanswer{ \tilde{\CA}(\omega)= \frac{\pi
e^{i\delta(\omega)}}{\sinh(\pi \omega/2)\sinh(\pi\omega)}. } We
notice that this amplitude is the same (up to an irrelevant
overall factor $e^{i\omega\infty}$) as the answer one would get by
replacing the integral in \owell\ by a sum of the residues at the
poles in $\rho(s)$ at $s=s_n\equiv\frac{i}{2}(2n+1)$, cf.
\McGreevyKB. The integrand in \owell, however, also has branch
cuts that (most likely) will spoil a precise correspondence with
the on-shell amplitude \newanswer. This is to be expected, since
the two descriptions of the tachyon condensation process, though
very similar, are in fact different.

Our adiabatic description \CAtimeint\ of the rolling tachyon has the
attractive property that we can ``see'' the tachyon condensing by
building up a growing open string tachyon potential on the
D-strings. Comparing with the matrix model calculation in
\McGreevyKB\ (see also \S4.2) furthermore makes the identification
between the boundary cosmological constant $\murb$ and the
eigenvalue coordinate $z$ manifest.

Taking into account the discussion from the previous subsection
we now also see what the final state is: The D-string has disappeared
altogether. This strongly suggests that there cannot be propagating
on-shell open strings in the final state, although they may have
been present in the initial state: There is no brane left to support
open strings for $t\ra\infty$. We also see what has happened to the
on-shell open strings that may have
been present in the initial state: They were pushed out to
$\varphi\ra-\infty$ during the process of tachyon condensation.

\subsec{Remarks on the Tachyon Bounce}

In this subsection, we will make a few somewhat speculative
comments about the open string spectrum on a D-string in the
presence of an open string tachyon that follows the bounce
trajectory.

To begin with, let us recall \LLM\ that the definition of the
one-point function $\langle\omega|B_{\lambda}\rangle^{}_{\rm\sst
bounce}$ is not unique and requires the choice of a contour. Two
natural choices of contour were discussed in \LLM, leading to the
results \eqn\xxxx{ \langle\omega|B_{\la}\rangle^{\rm\sst
HH}_{\rm\sst bounce}=\pi \frac{ \hat \lambda^{-i\omega}
}{\sinh(\pi\omega)}, \quad {\rm
and}\quad\langle\omega|B_{\la}\rangle^{\rm\sst real}_{\rm\sst
bounce}=2\pi \frac{ \sin(\omega\ln\hat
\lambda)}{\sinh(\pi\omega)}, } respectively, with $\hat \lambda
\equiv \sin \pi\lambda$. It would clearly be of interest to know
the spectrum of the $X^0$-CFT on the strip with boundary
conditions characterized by $|B_{\lambda}\rangle^{}_{\rm\sst
bounce}$. We have previously argued that there exists a continuous
part in the spectrum if the one-point function
$\langle\omega|B_{\la}\rangle^{}_{\rm\sst bounce}$ has a pole at
$\omega=0$. In this respect it is striking to observe that
$\langle\omega|B_{\la}\rangle^{\rm\sst HH}_{\rm\sst bounce}$ has
such a pole, whereas $\langle\omega|B_{\la}\rangle^{\rm\sst
real}_{\rm\sst bounce}$ does not have a pole at $\omega=0$.

A further hint about the spectrum of the $X^0$-CFT on the strip is
obtained by studying the analytic continuation of the partition
function of the corresponding Euclidean theory, which was
calculated in \refs{\PolchinskiMY,
\CallanUB}.
The result was \eqn\euclideancardy{ Z_{{\rm euc}}(g, q) = {1 \over
\eta(q)} ~ \int_{-\pi}^\pi {d\phi \over 2 \pi} ~ \sum_{n \in {\bf
Z}} q ^{ \left( n + \beta /4 \pi \right)^2 } \;, } where $\beta$
is related to $\phi$ via: $\sin(\beta/4) = \cos \left( \pi g
\right)~ \sin \left( \phi /2 \right)$.
We can rewrite \euclideancardy\ as
\eqn\eucpartitiontwo{
 Z_{\rm euc}(g,q) 
=
~\int\limits_{-\infty}^\infty\!\! d \sigma
\rho_{\rm euc}(\sigma) ~
{q ^{ (\sigma/4 \pi)^2}\over \eta(q)}\;
}
where the variable $\sigma$ is a continuation of
$\beta$ to the real line.
The spectral density is
\eqn\eucdensity{
 \rho_{\rm euc}(\sigma) = { d \phi \over d \sigma} =
{\rm \bf Re}~{1 \over 2} { |\cos \sigma/4| \over \sqrt{ \bar
\lambda^2 - \sin^2 \sigma/4}}
}
where $\bar \lambda \equiv \cos \pi g$.
As observed in \refs{\PolchinskiMY,\CallanUB}, the density of open-string
states has support in an infinite series of finite bands.
The bands connect into a single continuous spectrum at $g=0$,
when the boundary CFT represents a Neumann boundary condition,
and degenerate to a discrete spectrum for $g={1\over 2}$.

It seems natural to try and construct the partition function of
the corresponding Minkowskian theory by contour rotation. In order
to do this, one first of all has to represent \eucpartitiontwo\ as
a linear combination of integrals of \eucdensity\
over contours that pass the branch cuts in the upper or lower
half-plane. Naive continuation of \eucdensity\ to $\sigma/4 = i
\pi s$ yields
\eqn\plungingintothecomplexplane{
\rho(s) =
{ \pi \cosh \pi s \over \sqrt{ \sinh^2 \pi s -\bar \lambda^2}} }
where $\bar \lambda = \cos \pi \lambda$. It is tempting to view
$\rho(s)$ as the spectral density of the Minkowskian theory.
As a warning, however, we should add that one may have to face the
possibility that the final outcome for the Minkowskian partition
function critically depends on how one has chosen the contour.
Indeed we expect that the definition of the Minkowski boundary CFT
is not unique, but rather requires additional input. This is
natural from the point of view of the minisuperspace analysis of
\MaloneyCK, which suggests that the non-uniqueness in the
definition of the boundary CFT corresponding to $T_{\rm
bounce}(X^0)$ is directly related to the non-uniqueness of the
vacuum in general time-dependent backgrounds.\foot{In view of this
comment it is natural to ask why we did not encounter such
ambiguities in the case of $T_{\rm roll}(X^0)$. Our remark from
the end of the previous subsection offers an explanation: As the
D-string has disappeared after the tachyon has condensed, we can
not have outgoing on-shell open strings for $t\ra\infty$. This
corresponds to a preferred choice of boundary condition for the
Klein-Gordon operator in the mini-superspace analysis of
\MaloneyCK, and therefore to a preferred choice of the vacuum.
This offers an explanation for why the situation seemed to be much
clearer in the case of $T_{\rm roll}(X^0)$.}

In a later section, we will find an expression very similar to
\plungingintothecomplexplane\ from studying a classical rolling
eigenvalue in the matrix model. We expect that a comparison
between the CFT and matrix model result will provide useful
physical guidance in finding a precise definition of the Minkowski
theory in the presence of the tachyon bounce.

\subsec{Localized D-branes}

We will now briefly discuss the interpretation of the boundary
states proposed in \ZZB.
In the semiclassical limit, $b \to \infty$,
the interpretation of these boundary states is clear:
they describe branes
localized in the strong coupling region $\varphi\ra\infty$.
This can be seen from the fact that,
in this limit, the one-point function of the closed-string
tachyon on the disk with these boundary conditions
diverges at the boundary of the disk.

However, the point $b=1$ is in some sense quite far from the
semiclassical limit $b\ra 0$. We would therefore like
to convince ourselves that the main features of the
interpretation above persist for $b=1$.
To this aim let us consider $\langle \Psi_t|B_{\rm ZZ}\rangle$
for a time-dependent\foot{For the following Gedankenexperiment
we are not talking about 2d string theory, we just
consider the Liouville CFT.}
wave-packet
\eqn\xxx{
\langle\Psi_t|B_{\rm ZZ}\rangle=
\int_{0}^{\infty}\frac{dP}{2\pi}\;e^{-i\De_P t}
\bar{\Psi}(P)
\;\langle P|B_{\rm ZZ}\rangle,
}
where $\De_P=\frac{1}{4}(Q^2+4P^2)$. As discussed in some detail in \TB,
$\langle\Psi_t|$ describes a wave-packet that comes in
from $\varphi\ra -\infty$ for $t\ra -\infty$ and gets reflected
back into $\varphi\ra -\infty$ for $t\ra +\infty$. With the help
of the method of stationary phase it is easy to see that
the vanishing of the one-point function $\langle P|B_{\rm ZZ}\rangle$
at $P=0$ implies very rapid decay of $\langle\Psi_t|B_{\rm ZZ}\rangle$
for $t\ra\infty$. In other words, the coupling of closed strings
to $|B_{\rm ZZ}\rangle$ decays very rapidly for $\varphi \ra -\infty$.

On the other hand, we may note that the average energy $\bar{\De}_P$
of a wave-packet $\langle\Psi|$ gives a measure for how deeply
the packet penetrates into the strong coupling region. The one-point function
$\langle P|B_{\rm ZZ}\rangle$
diverges like $e^{\pi P}$ for $P\ra\infty$. This means that
$\langle\Psi|B_{\rm ZZ}\rangle$ grows exponentially if we
increase $\bar{\De}_P$. In other words, the deeper we probe into
the strong coupling region,
the stronger we ``feel'' the presence of the brane.

These localization properties, together with the fact that
the open string spectrum on the branes described with the help
of $|B_{\rm ZZ}\rangle$ is (almost) trivial makes
them natural candidates for the degrees of freedom that
define the holographic dual of the 2d closed string background.

\subsubsec{Closed-string emission from the rolling D-particle}

Consider the $(1,1)$ ZZ state
tensored with the bouncing boundary
state in the $X^0$ CFT, with the
Hartle-Hawking prescription.
The amplitude for emission of an on-shell closed string
from such a brane
is \LLM,\ZZB\
\eqn\notyetone{
\CA(\omega, P)
= \vev{ \omega | B_\lambda }_{\rm bounce}^{\rm HH}
\vev{ P | B_{ZZ}}
=  2 i \sqrt \pi \hat \lambda^{-i\omega}
~e^{ i \delta(2 P) }~ {\sinh 2\pi P
\over \sinh \pi \omega} ,
}
which upon setting $\omega = 2|P|$ becomes\foot{This
amplitude was considered
independently in \kms, where
in addition its interpretation in
the matrix model was understood.}
\eqn\one{
\CA(\omega)=
 \bigl[ \vev{ \omega | B_\lambda }_{\rm bounce}^{\rm HH}
\vev{ P | B_{ZZ}} \bigr]_{\omega = 2|P|}
= 2 i \sqrt \pi ~  e^{i\omega \ln \hat \lambda}
 ~e^{i \delta(\omega)}.
}
Other than the leg-pole factor, the wavefunction
of this source is a plane-wave in momentum space.
As a result, the eigenvalue-space
profile (obtained by stripping off
the leg-pole factor and fourier transforming)
is localized in space and time.
Reading off the time-delay from the phase factor,
the source is localized at the
point in spacetime where the eigenvalue
reaches the turning point.
This observation implies that we can think of
the ZZ state as a local source which
initializes the rolling trajectory.
\foot{
This observation leads us to the following point.
As pointed out in \McGreevyKB,
it is natural to associate
a Euclidean brane
with boundary tachyon
$$ T(\tilde X^0) = \hat \lambda \cos \tilde X^0$$
with the eigenvalue tunneling trajectory
with turning point at
$z = \hat \lambda$.
Tensoring this boundary state in the spacelike $\tilde X^0$ theory
with the D-particle state,
one finds a wavefunction in the
eigenvalue space which describes
exactly this process.}

\newsec{Rolling Eigenvalues: Classical Treatment}

In the next two sections we will study within the $c=1$ matrix
model, the creation of closed string modes due to an extra rolling
matrix eigenvalue. We will consider two situations: a classical
probe and a quantum probe. The classical probe is defined in
direct analogy with classical D-branes, and corresponds to an
extra eigenvalue that follows a prescribed classical trajectory.
As we will show in the next section, the quantum probe is obtained
from the classical probe by applying standard path-integral
quantization. Our goal is to understand the matrix model
calculations in terms of the CFT boundary states. We start with
the classical situation.

\subsec{Adding a Classical Eigenvalue}

Recall that the matrix quantum mechanics can be solved via a path-integral
(see \eg\ \GinspargIS),
by first discretizing time and starting from the (semi-infinite) matrix chain
model:
\eqn\eqmmb{\eqalign{Z_N &= \int\prod_{\alpha\leq q} {\rm d} M_\alpha\
e^{-\tr
(\sum_{\alpha\leq q} \bigl( V(M_\alpha)
-M_{\alpha-1} M_{\alpha}\bigr)}\cr
&=\int \prod_{{\scriptstyle \alpha\leq q}\atop {\scriptstyle i =1\dots N}}
\!\!{\rm d}\lambda_i^{(\alpha)}\ 
e^{-\sum_{i,\alpha}\bigl(V
(\lambda_i^{(\alpha)})-
\,\lambda_i^{(\alpha)}\lambda_{i}^{(\alpha+1)}\bigr)}
\Delta_{N}(\lambda^{(q)})\ .}}
where for the potential we take $V(M) = -\Tr M^2 + g\Tr M^4$ with $g$
very small; its dependence will drop out in the double scaled theory.
Upon taking the continuum limit and performing a Wick rotation, the above
discretized expression reduces to a path-integral over the trajectories
of the $N$ eigenvalues
\eqn\pathold{{Z_N(\tau)
= \int \prod_{i=1}^N
D \lambda_i(t)\, 
\, e^{{i\over \hbar} \int^\tau \! {\rm d} t \,
\sum_{i=1}^N\bigl(\dot \lambda_i^2 - V(
\lambda_i)\bigr)}\,\Delta_N(\tau)\ .}} This partition function
represents a quantum mechanical wave-function of the $N$
eigenvalues $\lambda_i = \lambda_i(\tau)$ at the time-instant
$\tau$ and satisfies the Schrodinger equation (here $p_i=
-i \hbar{\partial \over \partial \lambda_i(\tau)}$) \eqn\schr{i{\hbar} {d\
\over d\tau}Z_N(\tau) = \hat{H}_{old}(p,\lambda) Z_N(\tau)}
with\foot{Note that the power of the Vandermonde in this equation
is the inverse of what one would have naively expected by looking
at \pathold. The correct kinetic term of the Hamiltonian is
obtained by restricting the $U(N)$ invariant Laplacian to the
space eigenvalues. The difference arises due to the fact that the
canonical momentum $p_i$ and is not equal to the velocity $\dot
\lambda_i$ but related via $\dot\lambda_i= \Delta^{-1}p_i
\Delta$.} \eqn\ham{\hat{H}_{old}(p,\lambda)= \Delta_N^{-1}
\hat{H}_{N}
 \Delta_N,
~~~~~ \hat H_N = \sum_i \bigl(
{1 \over 4}p_i^2 + V(\lambda_i)\bigr).}
We wish to find a new Hamiltonian $H_{new}$ that represents the time-evolution
of the $N$ eigenvalues in the presence of an extra probe eigenvalue.
The way in which we will
do this is motivated by the way one treats D-particles
in string theory. Initially, the D-particle follows a classical trajectory,
\eqn\probeev
{\lambda_{N+1}(t) \equiv z(t). }
Eventually, to cancel string divergences, one is instructed to promote
$z$ to a quantum mechanical degree of freedom, and accordingly perform the path
integral over all possible trajectories $z(t)$,
weighted by an appropriate world-line action. After doing this
path-integral, the extra D-particle has become fully quantized and
indistinguishable from all other quantized D-particles. So our
definition of the wave-function $Z_{N}(\lambda;z(\tau))$ of the $N$ eigenvalues
in the presence of the probe $z$ must be such that:
\eqn\dbrane{Z_{N+1}(\tau) = \int \!\! Dz(t) \, e^{{i\over
\hbar} \int^\tau {\rm d} t \bigl(\dot z^2- V(
z) + \hbar \dot z A_B\!(z) \bigr)} \,
Z_{N} (\lambda\, ;\, z(t))\, .} Here we are including the possibility of a
Berry phase term in the classical probe action, to be specified below.
Its string theory interpretation is that the probe D-particle moves in
a non-trivial closed string tachyon background, and this may induce
such a term. Let us denote \eqn\berry{ e^{i \int^\tau \!\! {\rm d} t\,
\dot z A_B\!(z)} = e^{i \varphi_B(z)}}
The Berry connection is defined such that
\eqn\berrydef{\Bigl({\partial \over \partial z} + i A_B(z) \, \Bigr)\,
\tilde{Z}_{N}(\lambda; z) = 0}
with $Z_N(\lambda,z) = e^{-i\varphi_B} \tilde{Z}_N(\lambda,z)$.
The associated Berry phase is the
non-integrable (\ie\ path-dependent) phase factor. Together
with the interaction Hamiltonian $H_{int}$, it will ensure that,
upon quantization, the probe eigenvalue $z$ satisfies the proper Fermi
statistics with the other eigenvalues.

{}From its definition \dbrane, and using that $Z_{N+1}$ is as
given in \pathold\ with $N$ replaced by $N+1$, we find that the
new matrix wave-function $Z_N(\lambda,z)$ in the presence of the
probe eigenvalue $z$ is given by
\eqn\pathprobe{{Z_{N}(\lambda ; z)
= e^{-i{\varphi_B}(z)} \int \prod_{i=1}^{N}
D \lambda_i(t)\, e^{{i\over \hbar} \int^\tau {\rm d} t
\sum_{i=1}^{N}\bigl(\dot
\lambda_i^2- V( \lambda_i)\bigr)}\Delta_{N+1}(\tau)}\, .}
This new wavefunction satisfies the Schrodinger equation \eqn\schr{i\hbar
{d\ \over d\tau}Z_{N}(\lambda;z) = \hat{H}_{new}(p,\lambda) Z_N(\lambda;z)}
with \eqn\hamnew{\hat{H}_{new} = \Delta_{N+1}^{-1}\, \hat{H}_N\,
\Delta_{N+1}
}
where we used the property \berrydef\ of the Berry phase.
To extract the interaction Hamiltonian, we should compare the new and old
Schrodinger equations. We can write
\eqn\find{\hat{H}_{new} = \Delta_{N}^{-1}\, (\hat{H}_N + \hat{H}_{int})\,
\Delta_{N}} where (using that $\hbar$ is small) the
interaction Hamiltonian is given by
\eqn\hint{\hat{H}_{int}(z) = [\hat{H}_N,
\hat{\Phi}(z(t))]}
with
\eqn\loopz{\hat{\Phi}(z) 
= \log\Bigl({\Delta_{N+1}(z)\over \Delta_N}\Bigr)
= \sum_{i=1}^N \log(\lambda_i-z).}
\foot{
We take this opportunity to note
that the terms we have ignored in writing \hint\
are the matrix model description of the
contact terms involved in exponentiating
the boundary state into a shift
in the closed-string background
\GaiottoRM.
In this case, they have a very simple effect.
The expression
$ \hat H_N + \hat H_{\rm int} = e^{\hat \Phi} \hat H_N e^{-\hat \Phi} $
says that
$$ \hat H_{\rm int} = [ \hat \Phi, \hat H_N] + \half [ \hat \Phi, [\hat \Phi,
\hat H_N ] ]
+\dots.$$
But $ [ \hat \Phi, \hat H_N] = i \del_t \hat \Phi $ is the
field momentum of $\Phi$, and hence
the second term $ [ \hat \Phi, [\hat \Phi, \hat H_N]]$ is a
c-number, and as a result all successive terms vanish.
Thus, the effect is merely a multiplicative renormalization.
}
This is the interaction Hamiltonian used in \McGreevyKB.
The operator $\hat{\Phi}(z)$ is the Laplace transform of the
macroscopic loop operator in the $c=1$ matrix model. In
expectation values, it creates a hole in the large $N$ diagrams,
dual to the closed string worldsheets, with Dirichlet boundary
conditions in both the time and the eigenvalue direction.

\bigskip

\subsec{Bouncing Classical Eigenvalue as a D-brane}

In \McGreevyKB, the interaction Hamiltonian $H_{int}$ was used to
compute the particle production due to an extra classical rolling
eigenvalue $z(\tau) = \lambda e^\tau$. It was found that the
result corresponds with the closed string production due to a
rolling tachyon on the extended D-brane \FZZ, provided one uses
the adiabatic approximation discussed in \S3.3

This is a striking result for various reasons. First, while one
could hope that the eigenvalues have a direct relation with the
D-branes of the dual string theory, there exists no ironclad
reasoning that would guarantee a simple, direct correspondence
(see section 6). Secondly, one would normally expect
\PolyakovAF\poljoerg\kms\ that the eigenvalues should correspond
to the point- or particle-like D-brane states \ZZB. In the next
section, when we turn to the quantum mechanical probe, we will
reconcile our classical probe result with this viewpoint \kms.
First, we will present further evidence supporting the
identification of the classical probe trajectories with the
D-string boundary state.

Consider the one-parameter family of classical bounce trajectories
\eqn\pbounce{z(t) = \bar\lambda\sqrt{\mu} \cosh t} labelled by the
parameter $\bar{\lambda} \in [0,1]$. The limit $\bar\lambda
\downarrow 0$ (when defined appropriately\foot{ Specifically,
consider the trajectory $ z_{t_0}(t) = \bar{\lambda} \sqrt\mu
\cosh ( t - t_0) $, and fix $z_{t_0}(0) = \bar \lambda \cosh t_0 $
when taking $ \bar \lambda \to 0,~ t_0 \to \infty$. }) describes
the rolling trajectory $z(t) =\lambda e^t$ considered in
\McGreevyKB; the case $\bar\lambda =1$ describes an eigenvalue
that skims the Fermi sea.

In first order time-dependent perturbation theory, the emission
amplitude for closed string excitations is given by \eqn\ampli{
 \CA(\omega) =
\int \!\! dt\, e^{i\omega t} \,
\langle\mu_F+\omega| H_{int}|\mu_F\rangle
.}
The relevant matrix element of the macroscopic
loop operator takes the form \GinspargIS
\eqn\wzexp{
\langle\mu_F+\omega|
\hat{\Phi}(z(t))|\mu_F\rangle
=  e^{i\delta(\omega)} {\cos \pi \omega s(t) \over \
\omega \sinh \pi \omega}}
where
$s(t)$ parameterizes the probe trajectory via
\eqn\param{ z(t) = \sqrt{\mu} \, \cosh \pi s(t) }
The emission amplitude thus becomes
\eqn\ampli{
 \CA(\omega) =
\int \!\! dt\, e^{i\delta(\omega)-i\omega t} { \cos \pi \omega \,
s(t)\over \sinh  \pi \omega}.} This result must be compared with
the CFT prescription as given in \S3.3 of the bouncing tachyon
process on the D-string: the expression \ampli\ indeed precisely
matches with the CFT production amplitude \CAtimeint.

An immediate puzzle raised by this correspondence, however, is
that, as discussed in \S3.4, the D-string is expected to support a
continuous spectrum of open string excitations. (See however the
comment below eqn \xxxx.) Can we find a signal betraying the
presence of these open strings directly from the matrix model
amplitude \ampli? In particular, since the closed string one-point
amplitude is in fact an open string one-loop diagram, one may
expect an imaginary contribution coming from the possible pair
production of on-shell open strings.

\doublefig\realandimone{ As $t$ varies from zero to infinity, the
uniformizing variable $s$ takes a tour of the complex plane.
Depicted is the trajectory $ \cosh \pi s = 0.6 \cosh t $.
}{\epsfxsize2.0in\epsfbox{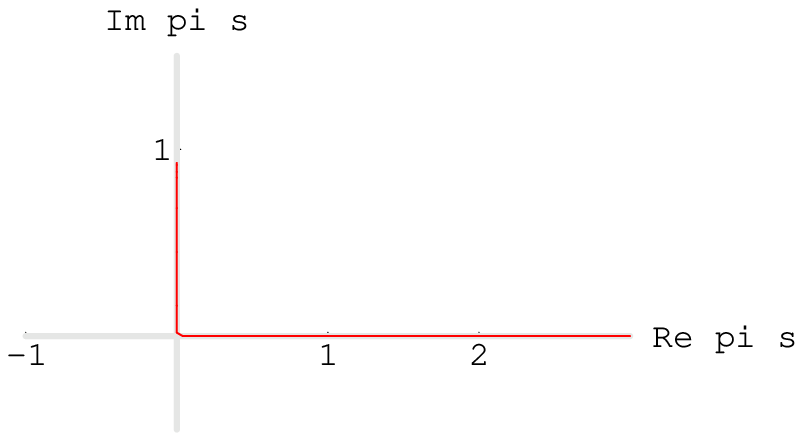}}
{\epsfxsize2.0in\epsfbox{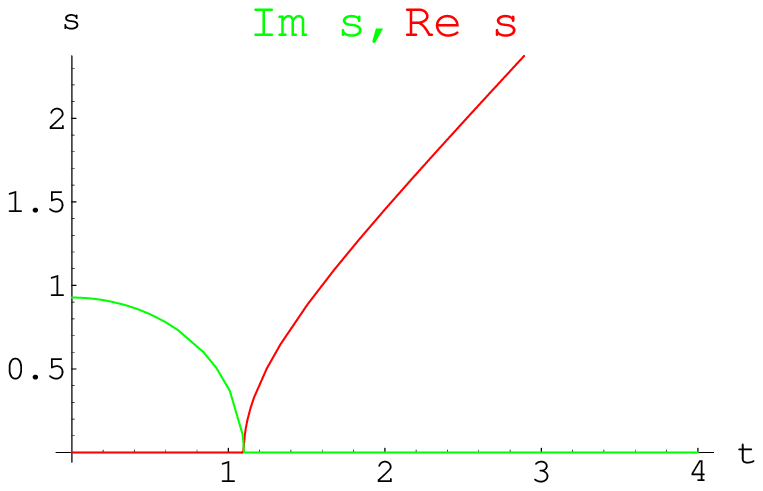}}

Let us look more closely at the time-dependence of the
uniformization parameter $s(t)$ that appears in \ampli. Combining
\param\ and \pbounce, we have \eqn\strel{\cosh \pi s(t) =
\hat\lambda \cosh t\, .} Clearly, as long as $\hat{\lambda} \cosh
t$ is larger than $1$, the parameter $s(t)$ is real. But there is
a short time interval (of order $\log \hat\lambda$ for small
$\lambda$) for which $\hat\lambda \cosh t$ is less than $1$, and
$s(t)$ becomes imaginary, see \realandimone.


We should note that the integrand in \ampli\ is purely
oscillatory, and the integral is therefore not unambiguously
defined. (The same remark applies of course to formula (3.10).)
Let us nevertheless proceed and change the variable of integration
from $t$ to $s$, with $t=t(s)$ defined by inverting \strel. We can
then write the amplitude \ampli\ as an integral over a judiciously
chosen contour in the complex $s$ plane, as follows
\eqn\probeelement{
{\cal A}(\omega) 
 =   \int \!\!
d s \, \rho(s) \,  e^{i\delta(\omega)-i\omega t(s)} {\cos\pi
\omega s\over \sinh \pi \omega}\, }
We now pick up the density \eqn\rhodefone{ \rho(s) = \; {dt\over
ds} \; \; = \; {\pi \sinh \pi s \over \sqrt{\cosh^2 \pi s -
\hat{\lambda}^2}}
} Here $\rho(s)$ can be thought of as a probability density
$|\psi(s)|^2$, since it is proportional to the amount of time the
eigenvalue spends at the location $s$. The above expression for
$\rho(s)$ is very similar to the naive formula
\plungingintothecomplexplane\ for the open string spectral density
for the tachyon bounce.  The two formulas differ, however, by an
interchange of $\cosh$ and $\sinh$-functions. Still, their formal
similarity is a suggestive piece of evidence relating the matrix
model to the continuum string theory, and may be of help in
finding the proper dictionary between the two. It will be
worthwhile to explore this particular correspondence further.






\newsec{Rolling Eigenvalues: Quantum Treatment }

The most convenient representation of the $c=1$ matrix model is as
a free 2-d fermion field theory \freeferm. In this section, we
will describe how, via standard path-integral quantization, our
classical probe eigenvalue can indeed be turned into a free
quantum mechanical fermionic particle.

Let us construct the free fermion field operator from the matrix
model. It is defined as the second quantized operator that creates
or destroys an eigenvalue: if we take its expectation value in the
$N\times N$ matrix integral as given in \pathold, it turns it into
an $N\!+\! 1 \times N\! +\! 1$ matrix integral (of the same form
\pathold\ but with $\Delta_N$ replaced by $\Delta_{N+1}$),
where it is understood that $\lambda_{N+1}(t)=z(t)$ starts (or
ends) its life at $t=t_0$. Following the same steps as presented
in the previous section, we can peel off the $N\times N$
expectation value from both sides, and obtain the following
operator identity: \eqn\relation{\Psi(z_0,t_0) = \!\!\!\!\!\!
\int\limits_{z(t_0)=z_0}\!\!\!\!\!\!\! Dz(t)\;
\e{\ {i\over \hbar} \int_{t_0} \!\! dt\, \bigl(\dot{z}^2 - V(z) +
\hbar A_B(z)\dot z + \hat{H}_{int}(z))}} with $A_B$ and $H_{int}$
as defined in \berrydef\ and \hint. Both sides of the above
equation should be read as quantum mechanical operators that act
on the $N\times N$ model defined by \pathold.

The equation \relation\ has a simple physical meaning: to describe the
quantum mechanical propagation of a particle, one can either use the
second quantized language and create the initial condition by means
of a local quantum field creating the particle, or one can sum over
all classical trajectories for the particle starting at a given point,
weighted by the classical action. In our context,
as we will see shortly, the equivalence between these two descriptions
will imply a very interesting relationship between the two types of
D-branes of 2-d string theory.

The formula \relation\ is clearly a bosonization/fermionization
formula. Let us make this more explicit. First we note that in
fact \eqn\aba{A_B(z) =
i \partial_z \hat{\Phi}(z)} which together with the form \hint\ of
$H_{int}$ allows us
to rewrite \relation\ as \eqn\bosonization{\eqalign{\Psi(z_0,t_0)
= \!\!\!\!\!\!
\int\limits_{z(t_0)=z_0}\!\!\!\!\!\!\! Dz(t)\;
\e{{i\over \hbar} \int_{t_0} \!\! dt\, \bigl(\dot{z}^2 - V(z) + i{d
\  \over dt}\, \hat\Phi(z)\, \bigr),}}} which can be further
simplified to \eqn\final{\Psi(z_0,t_0) = f(z_0,t_0) \, {\rm
exp}\bigl({{1\over \hbar} \hat\Phi(z_0,t_0)}\bigr) \, .} Here
$f(z_0,t_0)$ is a wave-function satisfying the single particle
Schrodinger equation with $H_0 = {1\over 4} p^2 + V(z)$, and the
second factor is the standard vertex operator expression for a
bosonized fermion.

At this point, let us recall
our discussion of classical and quantum D-branes in \S1.1 of the
Introduction. Comparing with the above results \relation\ and \final,
we see that the operator $\Psi(z,t)$ in
\final\ acts like the second quantized
field operator that creates the quantum version of the
D-string.  The form of the relation \final\ further suggests
that we should interpret the loop operator $\hat \Phi(z,t)$ as the
local quantum field obtained by bosonization
of the fermionic field $\Psi(z,t)$.
There is a subtlety in this relationship, however.
The standard collective field
$\phi(z,t)$ is related to the eigenvalue density
$\rho(z,t) = \tr \delta( z - Y(t)) $
by $ \rho(z,t)= 
\del_z \phi(z,t)$.
Therefore, comparing with the definition \loopz\
of the loop operator $\hat\Phi$, we see that
the collective field is obtained by
taking the discontinuity of $\hat\Phi$
across the branch cut,
\eqn\discontinuity{
2 \pi i \phi(z)
= \hat \Phi(z + i \epsilon) - \hat \Phi(z - i \epsilon),
}
turning the logarithm into a step function.
By applying this formula to
our first-quantized bosonization prescription,
we can thus extract
the local
fermionic operator $\psi(z,t)$ which
creates a quantum eigenvalue via
\eqn\oneeigenvalue{
 \psi(z,t) = e^{ 2 \pi i \phi(z,t)} .}
In this way, we recover the second quantized
picture of the creation of an eigenvalue used in \kms.

\subsubsec{A quantum descent relation}

The above results imply a remarkable relationship
when translated back into the language of 2-d string theory.

Recall that excitations of the collective
scalar field $\phi(z,t)$ correspond (up to
a semi-local field redefinition due to the legpole factor) to
excitations of the closed string tachyon. Following \kms, we can
go to the asymptotic region, where the eigenvalue motion becomes
relativistic and only depends on one light-cone coordinate $\tau$,
and expand the field in terms of tachyon creation operators
as
\eqn\expand{\phi(\tau) = \int \! {dp \over \sqrt {2\pi |p|}}
\; a^\dagger_p \; e^{-ip\tau}.} We can now make the comparison with the
on-shell emission amplitude \one\ due to a tachyon bounce on the
localized D-brane described in \S3.5. This emission amplitude is
given by  
\eqn\quantumdbrane{ \phi(\tau) = \int \! {dp \over 2\pi \sqrt
{2 |p|}}\; a^\dagger_p \; \langle p| B_{ZZ}\rangle\, \langle
\omega | B_\lambda\rangle_{\rm bounce}^{HH},} which indeed
exactly matches with the matrix model prediction, provided one
identifies $\tau =\log \hat \lambda$.
This identification was first obtained in \kms.

When combined with the relationship \discontinuity, this
identification leads to another interesting conclusion.  As we saw
before, the matrix element \wzexp\ of the loop operator coincides
with the boundary state wavefunction of the D-string.  The
D-string wavefunction indeed reveals the same analytic structure
when treated as a function of $z = \mu_B = \sqrt \mu \cosh \pi s
$: as pointed out in \emil\teleology\ it is multivalued on the
$\mu_B$ plane, and the discontinuity is equal to the wavefunction
of the D-particle \eqn\jump{ \ket{B_{ZZ}} = \ket{B_{s=2i} } -
\ket{ B_{s=0}}. } This equation should be compared with
\discontinuity, and can be viewed as a descent relation that views
the $s=2i$ D-string as a bound state of the $s=0$ D-string with
the D-particle. Thus things fit together nicely: the procedure
that extracts from the loop operator, that encodes the presence of
a classical probe eigenvalue, the local quantum field $\psi(z,t)$,
that creates and destroys eigenvalues, matches precisely with the
descent relation between the D-string and the D-particle
\teleology. This further supports the interpretation, put forward
in \kms, of $\psi(z,t)$ as the field that creates the ZZ
D-particle.


\bigskip

\newsec{Conclusion}

We would like to conclude this paper by addressing a few possibly
confusing points. All three issues revolve around the fact that
D-branes can dissolve into a closed string background.




\subsubsec{The raising of the Fermi sea}

The result \S5 seems to indicate that the fermion field operator
$\Psi(z_0,t_0)$, that creates the eigenvalue, can be completely
expanded in terms of perturbative closed string tachyon modes.
This assertion can be true only in the strict $\hbar$ to 0 limit.
For $\hbar$ finite, it is well-known that the fermion field and
corresponding vertex operator $e^{i\Phi}$ are {\it not} part of
the usual Fock space of the $\Phi$-field. $\Phi$ is a periodic
variable, and the total fermion number $Q_F = \int \! dz \,
\partial_z \Phi$ defines a superselection sector: no finite number
of bosonic oscillators can change $Q_F$. This statement is dual to
fact that in 2-d string theory, no finite number of normalizable
closed string tachyon modes can produce sufficient backreaction to
induce a shift in the non-normalizable zero-mode \cosmas.

\subsubsec{Worldsheet renormalization}

Based on experience and specific arguments
\ZamolodchikovGT\BanksQS\HughesBW\
one expects that the study of off-shell string theory requires
a violation of worldsheet conformal invariance.
Therefore, the reader may be given pause
by the adiabatic prescription of \S1.1, which
apparently provides a
prescription for studying arbitrary classical trajectories
of D-particles {\it using worldsheet conformal field theory.}

The procedure outlined in \S1.1
was inspired by the $c=1$ matrix model.
Indeed, when applied to two-dimensional string
theory it has produced results which we know
independently to be correct.

The resolution of this puzzle lies in the fact that
it is not those worldsheets whose conformal invariance
is violated --
this violation occurs on the worldsheets
made from the large-N diagrams of the open string theory.
These worldsheets clearly do not exhibit conformal
invariance before the double-scaling limit.
Taking the continuum limit implements
the approach to the IR fixed point.
One should be concerned, however,
about the application to
greater-than-two-dimensional models
where it will be difficult to realize
the higher-dimensional target space as
a continuum limit of the lattice theory.
In conclusion, the prescription we have
advocated {\it does} involve
non-conformal worldsheets,
but it is via the large-N RG that one
restores criticality.

\subsubsec{The identity of an eigenvalue}

By now, a number of quantitative matches have been made between
calculations involving branes in 2-d string theory and
calculations in the $c=1$ matrix model \McGreevyKB, \kms, \emil.
Inevitably, one encounters the question: ``Which brane {\it is}
the eigenvalue?''

In fact, this is not a
very well-posed question. Indeed, while both
types of D-branes appear to be very closely related to the
eigenvalues, neither is a perfect match: the classical extended
branes do seem to correspond to the classical eigenvalues, but
have many more degrees of freedom, while the localized branes
correspond to the quantum mechanical rather than classical
eigenvalues. There is also no reason to expect a perfect match:
the ${N+1}$-st eigenvalue will interact differently with the
background closed string theory defined by the collective field
theory of the first $N$ eigenvalues, than with the closed string
background provided by, say, the first $M$ eigenvalues. Following
the philosophy of \S6\ of \McGreevyKB, one could try to introduce
the notion of the ``bare'' string theory, which is the underlying
theory relative to which we define a decoupling limit of the
unstable D-particles, by sending $g_s \to 0$ while populating it
with $N\to \infty$ D-branes, keeping $\mu$ fixed. Since the
Liouville direction is {\it generated} by the matrices, the
``bare'' string theory knows nothing of it.

\bigskip
\centerline{\bf{Acknowledgements}}

\noindent
We would like to thank
Davide Gaiotto,
Nissan Itzhaki,
Igor Klebanov,
Juan Maldacena,
John Pearson,
Leonardo Rastelli, and Nathan Seiberg for discussions.
The work of JM is supported by a Princeton University Dicke Fellowship.
This work is supported by the National Science
Foundation under Grant No. 98-02484.
Any opinions, findings, and conclusions or recommendations expressed in
this material are those of the authors and do not necessarily reflect
the views of the National Science Foundation.

\listrefs
\end

\bye
\end

\eqn\finamp{
\CA_{\lambda}(P) =
c \int \!\! d\beta \, \rho(\beta) \,
e^{i\delta(P)}  {\cos \pi s P  \over \sinh
\pi P}}
We begin by considering the annulus amplitude
\eqn\ampl{
\CA_{\lambda}(P)=
\int\limits_0^1 \!\! d\tilde q\,
\bra{B_{\lambda}(\tau_1)}~\tilde q^{L_0}
| B_\lambda(\tau_0) \rangle  \langle\bra{V_P} ~\tilde q^{L_0}
| B_{s}\rangle
\cdot \CA_{\rm ghost} .}
Here $\langle\bra{V_P} $ is the Ishibashi
state built on the Liouville primary $V_P$.
The ghost part of the annulus is
\eqn\ghost{

Putting things together, we find that all factors $\eta(\tilde q)$ cancel,
leaving a trivial integration over the modular parameter $\tilde q$.
It reduces to the massless propagator
\eqn\pole{ \int\limits_0^1 \! {d \tilde q \over\tilde q} ~\tilde q^{P^2-\omega^2} = { 1
\over P^2-\omega^2} 
\; \longrightarrow \;
i \pi {\delta (P-\omega) \over \omega} ,}
so that we can use the delta-function to do the integral over $\omega$.
Collecting all the facts, we have
Now let us specialize the boundary states to fit the physical
problem we wish to study. First we set $s_1={i \over 2}$ so that
the state $\ket{B_{s_1}}$ represents the D-particle. The other
state we decompose as \be \bra{B_{s_2}} = \int \! {d\ell \over
\ell} \, e^{-\ell \sqrt{\mu} \cosh\pi s_2} \, \bra{W(\ell)}\, .
\ee The states $\bra{W(\ell)}$ we interpret as the position
eigenstate for the tachyon mode via the identification $\ell =
e^{\varphi}$. Using once again the integral formula ({\geez}),
we can thus write the amplitude as \be \CA_{s_1, s_2} = i c
\int_0^\infty {d\ell \over \ell}\, e^{- \ell \sqrt \mu \cosh \pi
s_2} { \pi K_{i\omega} (\sqrt \mu
\, \ell) 
\over \sinh (\pi \omega)\, \sinh(\pi \omega/2)}
\ee
The position-space profile
of the shift in the closed string tachyon is therefore
\be \delta {\cal T}(\ell, \omega) = c { \pi\, \ell^2 \,
K_{i\omega} (\sqrt \mu \, \ell) \over \sinh(\pi \omega)\,
\sinh(\pi\omega / 2) } \ee To extract the time-dependence of this
background, we Fourier transform using the contour in the figure;
the integral may be done by residues.

\figuur{contour}{8cm}{Fig 3.
The denominator
$$\sinh \pi \omega ~\sinh { \pi \omega \over 2}
= \pi^2 \prod_{n = 1}^\infty
\left( \omega^2 + (2n)^2\right)^2 \left(\omega^2 + ( 2n+1)^2 \right)
$$
has double zeros at even imaginary integers and
single zeros at odd imaginary integers.  The physical
tachyon response is obtained by Fourier transforming
using this contour.}
\be
\delta {\cal T}(\ell, t)
= \int {d\omega \over 2 \pi}\, e^{ i \omega t} \delta T(\ell, \omega)
= \delta {\cal T}(\ell) +
\sum_{ n = 1}^\infty
e^{- n t} c_n(\ell)
\ee
The terms with finite $n$ are transients which
represent the
splash of the probe D-particle into the Fermi sea.
The momenta of the transients are quantized in units of the
frequency of the harmonic oscillator
appearing in the euclidean continuation of the
matrix quantum mechanics.  The distinction between
odd and even multiples of this basic frequency
is the distinction between lengths and laps
under the barrier.

The static piece of the shift in the tachyon background is (with
$\ell = e^\varphi$) \be \delta {\cal T}(\varphi) \propto  \ell^2
\left.
\partial_\nu K_{\nu} ( \sqrt \mu \, \ell) \right|_{\nu = 0} =
\ell^2 I_0 (\sqrt \mu \ell). \ee which in the asymptotic region
amounts to a shift $\delta {\cal T}(\varphi) \propto e^{2
\varphi}$. Hence the shift at the location $\varphi =0$ of the
``dilaton wall'' is of order one. Since ${\cal T}(0) \simeq
{\mu\over 2} \log \mu$, we find that the presence of the extra
D-particle amounts to a shift $\delta \mu$ of order \be \delta \mu
\propto ( \log \mu)^{-1} , \ee in accordance with the
characteristic level density $\rho(\mu) = {\partial N\over
\partial \mu} \simeq -{2\over \pi} \log \mu$  of the $c=1$ matrix
model.



\begin{thebibliography}{99}

\lref\Sen:99}
A.~Sen, ``Non-BPS states and branes in string theory,''
arXiv:hep-th/9904207.


\lref\Sen:2002nu}
A.~Sen, ``Rolling tachyon,'' JHEP {\bf 0204}, 048 (2002)
[arXiv:hep-th/0203211];
``Tachyon matter,'' JHEP {\bf 0207}, 065 (2002)
[arXiv:hep-th/0203265];
``Field theory of tachyon matter,'' Mod.\ Phys.\ Lett.\ A {\bf
17}, 1797 (2002) [arXiv:hep-th/0204143].

\lref\Sen:2002vv}
A.~Sen, ``Time evolution in open string theory,'' JHEP {\bf 0210},
003 (2002) [arXiv:hep-th/0207105];
``Time and tachyon,'' arXiv:hep-th/0209122.

\lref\Gutperle:2002ai}
M.~Gutperle and A.~Strominger, ``Spacelike branes,'' JHEP {\bf
0204}, 018 (2002) [arXiv:hep-th/0202210]; A.~Strominger, ``Open
string creation by S-branes,'' arXiv:hep-th/0209090.

\lref\Gutperle:2003xf}
M.~Gutperle and A.~Strominger, ``Timelike boundary Liouville
theory,'' arXiv:hep-th/0301038.

\lref\stherm}
A.~Maloney, A.~Strominger and X.~Yin, ``S-brane thermodynamics,''
arXiv:hep-th/0302146.

\lref\Lambert:2003zr}
N.~Lambert, H.~Liu and J.~Maldacena, ``Closed strings from
decaying D-branes,'' arXiv:hep-th/0303139.

\lref\GIR}
D.~Gaiotto, N.~Itzhaki and L.~Rastelli, ``Closed Strings as
Imaginary D-branes,'' arXiv:hep-th/0304192.

\lref\longlist}
E.~Brezin, C.~Itzykson, G.~Parisi and J.~B.~Zuber,
``Planar Diagrams,''
Commun.\ Math.\ Phys.\  {\bf 59}, 35 (1978).

\lref\Ginsparg:is}
P.~Ginsparg
and G.~W.~Moore, ``Lectures On 2-D Gravity And 2-D String
Theory,'' arXiv:hep-th/9304011.

\lref\Klebanov:1991qa}
I.~R.~Klebanov, ``String theory in two-dimensions,''
arXiv:hep-th/9108019.

\lref\Polchinski:1994mb}
J.~Polchinski, ``What is string theory?,'' arXiv:hep-th/9411028.


\lref\matrix}
J. Warner and N. Warner, ``The Matrix: Reloaded,''
to appear.

\lref\Polchinski:1994jp}
J.~Polchinski, ``On the nonperturbative consistency of d = 2
string theory,'' Phys.\ Rev.\ Lett.\  {\bf 74}, 638 (1995)
[arXiv:hep-th/9409168].

\lref\Okuda:2002yd}
T.~Okuda and S.~Sugimoto, ``Coupling of rolling tachyon to closed
strings,'' Nucl.\ Phys.\ B {\bf 647}, 101 (2002)
[arXiv:hep-th/0208196].

\lref\Leblond:2003db}
F.~Leblond and A.~W.~Peet,
arXiv:hep-th/0303035.


\lref\kutasov:2003er{
D.~Kutasov and V.~Niarchos,
``Tachyon effective actions in open string theory,''
arXiv:hep-th/0304045.

\lref\Polchinski:mf{
J.~Polchinski,
``Critical Behavior Of Random Surfaces In One-Dimension,''
Nucl.\ Phys.\ B {\bf 346}, 253 (1990).

\lref\Seiberg:1990eb{
N.~Seiberg,
``Notes On Quantum Liouville Theory And Quantum Gravity,''
Prog.\ Theor.\ Phys.\ Suppl.\  {\bf 102}, 319 (1990).

\lref\Polchinski:1990mh{
J.~Polchinski,
``Remarks On The Liouville Field Theory,''
UTTG-19-90
{\it Presented at Strings '90 Conf., College Station, TX, Mar 12-17, 1990}


\lref\Fateev:2000ik{
V.~Fateev, A.~B.~Zamolodchikov and A.~B.~Zamolodchikov,
``Boundary Liouville field theory. I: Boundary state and boundary  two-point function,''
arXiv:hep-th/0001012.}

\lref\Teschner:2000md{
J.~Teschner,
``Remarks on Liouville theory with boundary,''
arXiv:hep-th/0009138.

\lref\Rajaraman:1999hn{
A.~Rajaraman and M.~Rozali,
``D-branes in linear dilaton backgrounds,''
JHEP {\bf 9912}, 005 (1999)
[arXiv:hep-th/9909017].

\lref\Ponsot:2001ng{
B.~Ponsot and J.~Teschner,
``Boundary Liouville field theory: Boundary three point function,''
Nucl.\ Phys.\ B {\bf 622}, 309 (2002)
[arXiv:hep-th/0110244].

\lref\Kostov:2002uq{
I.~K.~Kostov,
``Boundary correlators in 2D quantum gravity: Liouville versus discrete  approach,''
arXiv:hep-th/0212194.}

\lref\Moore:1991ir{
G.~W.~Moore, N.~Seiberg and M.~Staudacher,
``From loops to states in 2-D quantum gravity,''
Nucl.\ Phys.\ B {\bf 362}, 665 (1991).}

\lref\Kazakov:1991pt{
V.~A.~Kazakov and I.~K.~Kostov,
``Loop gas model for open strings,''
Nucl.\ Phys.\ B {\bf 386}, 520 (1992)
[arXiv:hep-th/9205059].}

\lref\dj{
S.~R.~Das and A.~Jevicki, ``String Field Theory And Physical
Interpretation Of D = 1 Strings,'' Mod.\ Phys.\ Lett.\ A {\bf 5},
1639 (1990).

\lref\Polchinski:fq{
J.~Polchinski,
``Combinatorics Of Boundaries In String Theory,''
Phys.\ Rev.\ D {\bf 50}, 6041 (1994)
[arXiv:hep-th/9407031].}


\bye